\documentclass[12pt]{article}

\usepackage{newtxtext,newtxmath}
\usepackage{amsmath, bm}
\usepackage{graphicx}
\usepackage[utf8]{inputenc}
\DeclareUnicodeCharacter{2009}{\,}
\usepackage[letterpaper,margin=1in]{geometry}
\renewenvironment{abstract}
	{\quotation}
	{\endquotation}

\date{}

\makeatletter
\renewcommand{\fnum@figure}{\textbf{Figure \thefigure}}
\renewcommand{\fnum@table}{\textbf{Table \thetable}}
\makeatother

\usepackage{scicite}

\usepackage{url}

\def\scititle{

Observation of OAM non-conservation in entangled photon generation }

\title{\bfseries \boldmath \scititle}

\author{
	Suman Karan$^{1\ast}$, and
	Anand K. Jha$^{1\ast}$\and
	\small$^{1}$Department of Physics, Indian Institute of Technology Kanpur, Kanpur, UP 208016, India.\and
	\small$^\ast$Corresponding author. Email: sumankaran2@gmail.com, akjha@iitk.ac.in
}

\begin{document} 

% Insert the title and author list
\maketitle

\begin{abstract} \bfseries \boldmath

Orbital angular momentum (OAM)-entangled states produced by spontaneous parametric down-conversion (SPDC) are considered ideal for realizing high-dimensional entangled states, which have several advantages for quantum technologies. However, the limited sensitivity of current two-photon OAM detectors is a major roadblock not only for realizing such technologies but also for resolving foundational questions, such as OAM conservation in SPDC. The current theoretical understanding is that OAM is not conserved in Type-II SPDC but is conserved in Type-I. Experimentally, although non-conservation in Type-II has not been demonstrated, conservation in Type-I has been reported frequently and has become an underlying assumption for techniques generating high-dimensional OAM entangled states. In this work, we experimentally demonstrate a high-sensitivity two-photon OAM detector, using which, contrary to the current understanding, we report non-conservation of OAM in Type-I SPDC. We attribute this to a spatial walk-off effect and prove it using a framework free of standard phase-matching approximations.

\end{abstract}

% The first paragraph of any Science paper does NOT have a heading
% Nor is it indented

\pagebreak
\noindent

The high-dimensional basis of orbital angular momentum (OAM) offers significant advantages for photonic quantum technologies compared to the two-dimensional polarization basis. These include terabit-scale data transmission using OAM-mode multiplexing for long distance communication  \cite{wang2012natphoton, bozinovic2013science}, enhanced security and error tolerance for quantum communication protocols \cite{cerf2002prl,nikolopoulos2006pra}, efficient gate implementation \cite{ralph2007pra,lanyon2009naturephys}, super-sensitive measurement in quantum metrology \cite{dambrosio2013natcomm, jha2011pra}, and fundamental tests of quantum mechanics \cite{dada2011natphy, law2004prl, miatto2012epjd}. The most-widely used process for generating high-dimensional OAM entangled two-photon state is spontaneous parametric down-conversion (SPDC)—a second-order nonlinear process in which a pump photon at higher frequency down-converts into two photons of lower frequencies, referred to as the signal and idler photons \cite{klyshko2018, boyd2008}. Despite numerous protocols for harnessing OAM entanglement being around, the practical implementations have remained a challenge due to the lack of efficient high-sensitivity detectors for measuring the OAM of entangled photons.

The first major advancement in this direction was the development of a technique for measuring OAM modes and their superpositions using a spatial light modulator (SLM) and a single-mode fiber (SMF) \cite{mair2001nature}. Even currently, it is the most widely used technique for measuring OAM and has facilitated experimental investigations of some of the foundational questions, in particular, the conservation of OAM in SPDC. OAM conservation in SPDC implies that the sum of the OAMs of the down-converted signal and idler photons equals the OAM of the pump photon. Prior to the development of the above technique, a few experimental works had hinted at non-conservation of OAM in SPDC \cite{arlt1999pra, arnaut2000prl}, but these results were mostly speculative. However, with the above detection scheme, OAM conservation in Type-I SPDC was demonstrated directly for the first time. Although no rigorous theoretical proof supported this experimental demonstration, it led to the widespread acceptance that OAM remains conserved in type-I SPDC, which, since then has become the underlying assumption of techniques for generating high-dimensional OAM-entangled state \cite{karan2023prapplied, vaziri2003prl, wang2017optica, liu2020sciadvances, qiu2023natcomm, zhang2017natcomm} as well as of applications in quantum communication \cite{bechmannpasquinucci2000prl, nikolopoulos2006pra, cerf2002prl, ecker2019prx, sit2017optica, bouchard2018oe}, quantum computing \cite{hu2018sciadvances}, quantum imaging \cite{chen2014lsa}, and fundamental tests of quantum mechanics \cite{dada2011natphy, law2004prl, miatto2012epjd}. Although a subsequent theory work \cite{frankearnold2002pra} supported the observations of Ref.~\cite{mair2001nature}, it involved several phase-matching approximations. A few years later, OAM conservation was demonstrated even in Type-II \cite{walborn2004pra}; however, a detailed theoretical analyses pointed out that OAM is conserved only in Type-I but not it Type-II \cite{barbosa2007pra, feng2008prl}. The physical reasoning for this was as follows: in Type-II phase matching, one of the entangled photons is extraordinary-polarized and experiences transverse walk-off in propagating through the anisotropic nonlinear crystal, whereas the other photon is ordinary-polarized and faces no walk-off \cite{barbosa2007pra, feng2008prl}, and this imbalance causes the non-conservation \cite{ rubin1996pra}. However, in Type-I SPDC, since both the photons are ordinary-polarized there is no such imbalance and the OAM remains conserved \cite{barbosa2007pra, feng2008prl}. In addition to Type-II SPDC, there have been calculations showing that non-conservation of OAM takes place in extremely non-collinear Type-I SPDC  as well \cite{molinaterriza2003optcomm}. Nonetheless, till date, there has been no experimental demonstration of non-conservation in either Type-I or Type-II SPDC.

Thus the current understanding on OAM conservation relies on theoretical models that employ severe phase-matching approximations and the two-photon OAM measurement techniques that have limited detection sensitivities. It is now known that the widely-used detection scheme of Ref.~\cite{mair2001nature} has mode-dependent detection efficiency, and it post-selectively measures only the lowest order radial mode, resulting in an OAM spectrum that is not the true spectrum \cite{qassim2014josab}. Several other OAM measurement techniques have also been developed but none comes close to measuring OAM with measurement fidelities high enough to resolve the conservation question. For example, the angular-slit based techniques involve significant photon loss \cite{jha2010prl, jha2011pra, malik2012pra}. The interferometric methods have been demonstrated only for pure states \cite{kulkarni2020prapplied, malik2014natcomm} and diagonal mixed states \cite{kulkarni2017natcomm, offer2018commphy, pires2010prl, peeters2007pra} and thus require prior knowledge about the state. The camera-based coincidence techniques  have been demonstrated only for pure states and suffer from pixel resolution effects \cite{zia2023natphoton, dehghan2024optica}. Hybrid approaches combining ICCD cameras with single-mode fibers have their own set of limitations \cite{ibarraBorja2019oe, li2023prl}, stimulated emission tomography suffers from pixelation artifacts \cite{xu2024prresearch}, and  recently-demonstrated broadband uniform-efficiency detectors work only for single-photon states \cite{karan2025sciadv}. 

In contrast, in this article, we report a technique for measuring the OAM spectrum of entangled photons that has uniform detection efficiency over a broad range of modes, does not involve any post-selection or loss, does not rely on any prior information about the state, has very high signal-to-noise ratio, and is able to measure the true joint OAM spectrum of entangled photons. Using this detector, which is based on measuring the angular correlation function of entangled two-photon field, we experimentally demonstrate non-conservation of OAM in Type-I SPDC. In order to quantify the effect, we construct a non-conservation parameter $\mathcal{N}$ that ranges from 0 to 100$\%$, with $\mathcal{N}=0$ representing perfect OAM conservation. We report experimental observation of  $\mathcal{N}=42.92\%$ with a 15-$mm$ thick nonlinear crystal in collinear Type-I phase-matching condition. By using a theoretical framework free from standard phase-matching approximations, we are able to numerically model the experimental observations with up to 87$\%$ fidelity.  Our work could be an important step towards more accurate and sensitive measurement of the OAM of entangled photons, and we therefore expect it to have important consequences for OAM based high-dimensional quantum technologies.

%breaks rotational symmetry in type-I SPDC, with $\mathcal{N}$ scaling quantitatively with $\alpha_p L$. When $\alpha_p L=0$, OAM is conserved, confirming that spatial walk-off of the pump causes the symmetry breaking. 
  
%These findings extend beyond SPDC to any nonlinear anisotropic media. The $\alpha_p L$ scaling provides a predictive framework for angular momentum transfer in other nonlinear processes such as sum-frequency generation, four-wave mixing, and optical parametric amplification, where similar walk-off should break rotational symmetry and lead to non-conservation of angular momentum. For quantum technologies employing type-I SPDC for high-dimensional OAM encoding—including quantum communication, quantum computing, and quantum metrology—this non-conservation requires reanalysis of protocols assuming perfect conservation and enables optimization through crystal parameter selection. These results fundamentally revise our understanding of entangled photon generation in anisotropic nonlinear media and resolve the long-standing question of OAM conservation in SPDC.

\section*{Concept}
\begin{figure}[t!]
\centering
\includegraphics[scale=1]{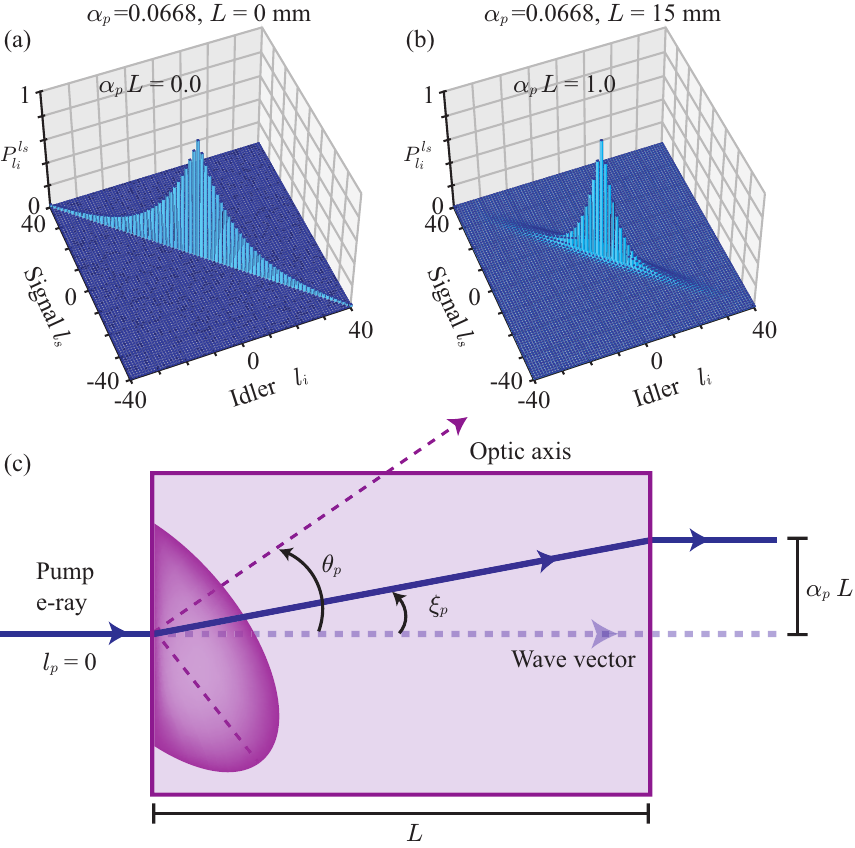}
\caption{{\bf Non-conservation of OAM in SPDC.} (a) and (b) are the numerical plots of $P^{l_s}_{l_i}$ for $L=0$ mm and $L=15$ mm, respectively. (c) A conceptual depiction of non-conservation of OAM in Type-I SPDC. An extraordinary-polarized pump photon propagating through an anisotropic crystal experiences spatial walk-off by $\alpha_p L$, where $\alpha_p = \tan \zeta_p $ and $L$ is the crystal thickness; $\theta_p$ denotes the angle between the pump wave vector direction and the crystal optic axis.  When $\alpha_p L \neq 0$, spatial walk-off induces a spread in the OAM distribution of the idler photon resulting in non-conservation on OAM in Type-I SPDC.}
\label{fig:conceptual_fig_non_conservation}
\end{figure}

\noindent{\bf The state of the OAM-entangled photons.} Photons in the Laguerre-Gaussian (LG) spatial modes $LG^{|l|}_p\left(\rho, \phi\right)= LG^{|l|}_p\left(\rho\right)e^{-il \phi}$ carry orbital angular momentum (OAM) of $l \hbar$, where $l$ is the OAM mode index which can take any integer value and $p$ is the radial mode index  \cite{allen1992pra}. The state $|\psi_2\rangle$ of the SPDC photons in the Laguerre-Gaussian (LG) basis can be represented as \cite{torres2003pra, miatto2011pra, jha2011pra}: 
\begin{align}
|\psi_2\rangle=\sum_{l_{s}}\sum_{l_{i}}\sum_{p_s}\sum_{p_i} C^{l_{s},p_{s}}_{l_i, p_i} |l_{s}, p_s\rangle_s|l_{i}, p_i\rangle_i \label{two-photon state}
\end{align} 
Here, the subscripts $s$ and $i$ stand for signal and idler, respectively, and $|l_{s}, p_s\rangle_s$ represents the state of the signal photon in the Laguerre-Gaussian (LG) basis defined by the OAM-mode index $l_{s}$ and the radial index $p_s$, etc. The complex coefficient $C^{l_s,p_s}_{l_i, p_i}$ is given by \cite{kulkarni2018pra}: $ C^{l_{s},p_{s}}_{l_i,p_i}=\!\! \iint \!\! V({\bm q_s}, {\bm q_i})\Phi({\bm q_s}, {\bm q_i})LG^{*l_{s}}_{p_s}({\bm q_s})LG^{*l_{i}}_{p_i}({\bm q_i})d{\bm q_s} d{\bm q_i}$, where ${\bm q_s}\equiv(q_{sx}, q_{sy})=(\rho_s\cos\phi_s,  \rho_s\sin\phi_s)$, ${\bm q_i}\equiv(q_{ix}, q_{iy})=(\rho_i\cos\phi_i,  \rho_i\sin\phi_i)$, $d{\bm q_s}=\rho_s d\rho_s d\phi_s$, and $d{\bm q_i}=\rho_i d\rho_i d\phi_i$. Here, $V({\bm q_s}, {\bm q_i})$ is the pump field amplitude, $\Phi({\bm q_s}, {\bm q_i})$ is the phase matching function  and $LG^{l_s}_{p_s}({\bm q_s})$ is the momentum-basis representation of state $|l_s, p_s \rangle_s$, etc.  The probability that the signal and idler photons are detected with OAMs $l_s\hbar$ and $l_i\hbar$, respectively, is calculated by summing over radial indices, that is, $P_{l_s}^{l_i} = \sum^{\infty}_{p_s,p_i =0}\langle|C_{l_i,p_i}^{l_s,p_s}|^2\rangle_e$, and it can be shown to be \cite{kulkarni2018pra, karan2020joopt}:
\begin{align}
P^{l_s}_{l_i} =\frac{1}{4\pi^2}\iint_{0}^{\infty} {\Big |}\iint_{-\pi}^{\pi} V(\rho_s, \rho_i,\phi_s,\phi_i) \Phi(\rho_s, \rho_i, \phi_s,\phi_i) 
e^{i(l_s\phi_s+l_i\phi_i)}d\phi_s d\phi_i  {\Big |}^2 
\rho_s\rho_i d\rho_s d\rho_i. \label{eqn:spdc_p_ls_li}
\end{align}
Eqs.~(\ref{two-photon state}) and (\ref{eqn:spdc_p_ls_li}) are the most general expressions for the state and the joint probability. However, in most experiments, involving a Gaussian pump field ($l_p=0$), the state of the SPDC photons is taken as the Schmidt-decomposed OAM-entangled state \cite{law2004prl, torres2003pra, pires2010prl, dada2011natphy, miatto2011pra, jha2011pra, mair2001nature}:
\begin{align}
|\psi_2\rangle=\sum_{l=-\infty}^{\infty}\sqrt{S_l}|l\rangle_s|-l\rangle_i, \label{two-photon state-Schmidt}
\end{align}
where the Schmidt spectrum $S_l= P_l^{-l}$ is the probability that the signal and idler photons have OAMs $l\hbar$ and $-l\hbar$, respectively. The above form yields the maximally entangled state in an $N$-dimensional sub-space if $S_l=\frac{1}{\sqrt{N}}$ $\forall N$. The fact that Eq.~(\ref{two-photon state-Schmidt}) cannot be derived from Eq.~(\ref{two-photon state}) can be easily seen by noting that summing over radial indices should yield a mixed two-photon state and not a pure two-photon state as in Eq.~(\ref{two-photon state-Schmidt}). Nonetheless, Ref.~\cite{mair2001nature} experimentally demonstrated that for a Gaussian pump field ($l_p=0$), whenever a signal photon is found to have the OAM mode index $l$, the idler photon is guaranteed to have $-l$ as the OAM mode index. This result was taken to imply OAM conservation as $l_p=l_s+l_i$. And since then, Eq.~(\ref{two-photon state-Schmidt}) is taken as the state of the SPDC photons.

For a Gaussian pump beam, $V(\rho_s, \rho_i,\phi_s,\phi_i) = {\rm exp}\left\lbrace\frac{w_o^2}{4}\left[\rho^2_s + \rho^2_i + 2\rho_s \rho_i \cos\left(\phi_s - \phi_i\right)\right]\right\rbrace$, and $\Phi(\rho_s, \rho_i,\phi_s,\phi_i) = L {\rm Sinc}\left[\Delta K_z(\rho_s, \rho_i,\phi_s,\phi_i) \frac{L}{2}\right] {\rm exp}\left\lbrace i \Delta K_z(\rho_s, \rho_i,\phi_s,\phi_i) \frac{L}{2} \right\rbrace$. And for type-I degenerate SPDC in a BBO crystal, $\Delta K_z(\rho_s, \rho_i,\phi_s,\phi_i)$ can be expressed as 
\begin{align}\label{eqn:delkz}
&\Delta K_z(\rho_s, \rho_i,\phi_s,\phi_i)= k_{sz} + k_{iz} -  k_{pz} = k_{po}\left(n_{so} - \eta_p\right) - \dfrac{1}{n_{so} k_{po}}\left(\rho^2_s + \rho^2_i\right) + \alpha_p\left[\rho_s \cos \phi_s + \rho_i \cos \phi_i \right] \nonumber \\
& \qquad\qquad\qquad\qquad+ \dfrac{1}{2 \eta_p k_{po}}\left[ \beta^2_p\left(\rho_s \cos \phi_s + \rho_i \cos \phi_i\right)^2 + \gamma^2_p\left(\rho_s \sin \phi_s + \rho_i \sin \phi_i \right)^2\right],
\end{align} 
where
\begin{align} \label{eq: alphap_phasematch}
&\alpha_p = \dfrac{(n^2_{po}- n^2_{pe})\sin\theta_p \cos\theta_p}{n^2_{po}\sin^2\theta_p + n^2_{pe} \cos^2 \theta_p}, \qquad \beta_p = \dfrac{n_{po} n_{pe}}{n^2_{po}\sin^2\theta_p + n^2_{pe} \cos^2 \theta_p},\\
&\gamma_p = \dfrac{n_{po} }{\sqrt[]{n^2_{po}\sin^2\theta_p + n^2_{pe} \cos^2 \theta_p}}, \qquad \eta_p = \dfrac{n_{po} n_{pe} }{\sqrt[]{n^2_{po}\sin^2\theta_p + n^2_{pe} \cos^2 \theta_p}}. \nonumber
\end{align}
Here $n_{po}$, $n_{pe}$ are the refractive indices of the ordinary and extraordinary polarizations at the pump wavelength. $n_{so}$ is the refractive index of the ordinary polarization at the degenerate signal and idler wavelength, $L$ is the crystal thickness, and $\theta_p$ is the phase-matching angle---an angle between the crystal optic axis and the pump propagation direction. 

\begin{figure}[hbtp]
\centering
\includegraphics[scale=0.94]{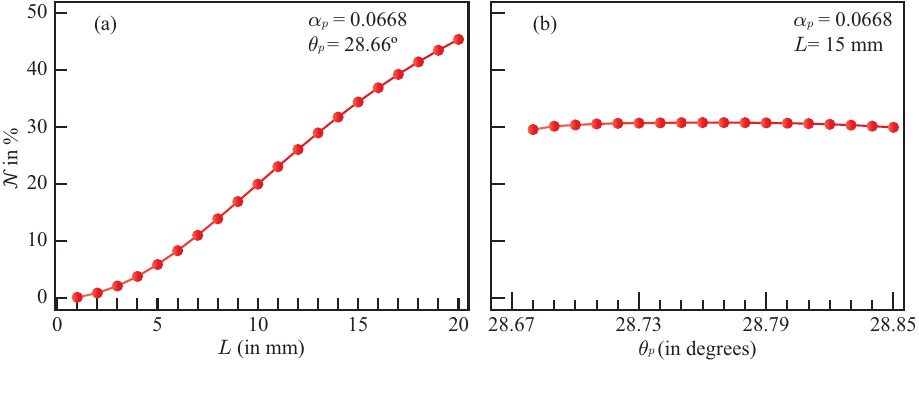}
\caption{{\bf Effect of crystal thickness $L$ and phase matching angle $\theta_p$ on OAM non-conservation.} (a) Plot of $\mathcal{N}$ as a function of $L$ at fixed phase matching angle $\theta_p = 28.66^{\circ}$. (b) Plot of $\mathcal{N}$ as a function of $\theta_p$ at fixed crystal thickness $L=15$ mm. For both the plots, $\alpha_p = 0.0668$.}
\label{fig:nl_vs_alpha_l_thetap}
\end{figure}

\noindent{\bf Non-conservation of OAM in SPDC.} Figure \ref{fig:conceptual_fig_non_conservation} (a) and \ref{fig:conceptual_fig_non_conservation}(b) show numerically evaluated $P^{l_s}_{l_i} $ for $L=0$ mm and $L=15$ mm, respectively. For both the plots, the pump field is Gaussian ($l_p = 0$) with $w_0=388~\mu$m, $\theta_p=28.66^{\circ}$. We find that with $L=0$, the OAM is conserved since $P^{l_s}_{l_i}=0$ for $l_s \neq l_i$. However, for $L=15$ mm, $P^{l_s}_{l_i}\neq 0$ for $l_s \neq l_i$, implying non-conservation of OAM in Type-I SPDC. This result is contrary to the current understanding, which is that since in Type-I SPDC the signal and idler photons do not experience any spatial walk-off, the OAM should remain conserved even if it is not conserved in Type-II SPDC \cite{barbosa2007pra, feng2008prl}. However, we find that this understanding overlooks the fact that in type-I SPDC, the pump beam is extraordinary-polarized \cite{karan2020joopt, rubin1996pra} which experiences walk-off caused by finite $\alpha_p L$. When $\alpha_p L$ approaches zero, there is no walk-off and the OAM remains conserved. However, when $\alpha_p L \neq 0$, there is an effective spatial walk-off resulting in non-conservation. For conceptual clarity, we illustrate this in Fig.~\ref{fig:conceptual_fig_non_conservation} (c). The extraordinary-polarized pump propagates through the anisotropic crystal at angle $\theta_p$ relative to the crystal optic axis, causing its Poynting vector to deviate by angle $\zeta_p$ from the wave vector direction due to crystal's anisotropy. The figure also shows the dispersion ellipse for the negative uniaxial BBO crystal. As the pump beam propagates through the crystal of thickness $L$, the angular deviation $\zeta_p$ results in a lateral displacement $\alpha_p L$, where $\alpha_p = \tan\zeta_p$ (see Supplementary Materials section~\ref{sec:si_alphap_zetap} for the derivation). The walk-off shifts the center of the pump beam causing appearance of newer OAM modes around the original beam center, leading to non-conservation of OAM.

%perfect OAM conservation when $\alpha_p L = 0$ requires that the generated signal and idler photons satisfy $l_p = l_s + l_i$. In this case, detecting the signal photon at $l_s = 0$ would yield simultaneous detection of an idler photon sharply peaked at $l_i = 0$ with no spectral broadening. However, when $\alpha_p L \neq 0$, detecting the signal photon at $l_s = 0$ yields simultaneous detection of an idler photon OAM distribution centered at $l_i = 0$ but with finite spectral width. Therefore, $l_p \neq l_s + l_i$, demonstrating OAM non-conservation despite both signal and idler photons being ordinary-polarized in type-I phase matching.

\noindent{\bf Quantifying the non-conservation of OAM in SPDC}: To quantify the degree of non-conservation, we define a non-conservation parameter $\mathcal{N}$ as:
\begin{equation}\label{eqn:n_l expression}
\mathcal{N}  =\left[1- \dfrac{\sum_{l_i=-l_s}p^{l_s}_{l_i} }{\sum_{l_s,l_i} p^{l_s}_{l_i}}\right]\times 100\%,
\end{equation}
where $\mathcal{N} = 0\%$ represents perfect conservation and  $\mathcal{N} = 100\%$ represents maximum non-conservation, which refers to the situation in which $P^{l}_{-l} = 0$ $\forall l$. Fig.~\ref{fig:nl_vs_alpha_l_thetap}(a) presents numerically evaluated $\mathcal{N}$ as a function of $L$ for the phase matching angle $\theta_p = 28.66^{\circ}$. We find that for $L=0$, $\alpha_p L=0$, implying OAM conservation, that is, $\mathcal{N} = 0\%$. As $L$ increases, $\alpha_p L$ becomes more prominent, resulting in greater non-conservation, with up to $\mathcal{N}=50\%$ for $L=20$ mm. The effect of $\theta_p$ on $\mathcal{N}$ is shown in Fig.~\ref{fig:nl_vs_alpha_l_thetap}(b) for a BBO crystal of thickness $L=15$ mm. We observe that varying $\theta_p$ does not have significant effect on $\mathcal{N}$. Thus, contrary to theoretical works predicting non-conservation at extremely non-collinear phase-matching  \cite{molinaterriza2003optcomm}, we find that phase-matching does not much affect OAM non-conservation.

The widespread acceptance of OAM conservation following observations of Ref.~\cite{mair2001nature} can be attributed to two factors. First, the OAM detection technique introduced in Ref.~\cite{mair2001nature} does not measure the true OAM spectrum. This technique is sensitive only to the $p = 0$ mode of the field \cite{agrawal_springer, karan2023prapplied, qassim2014josab}. However, SPDC-generated entangled states contain significant contributions from $p\neq 0$ modes, which remain completely undetected in this technique \cite{miatto2011pra, zhang2014pra}. Second, although the effect of non-conservation increases with $L$, Ref.~\cite{mair2001nature} had used relatively thinner crystals, that is, $L=2$ mm, resulting in smaller non-conservation effect, which could not be unambiguously detected due to the low overall detection sensitivity of the technique.

\begin{figure}[hbtp]
\centering
\includegraphics[scale=0.94]{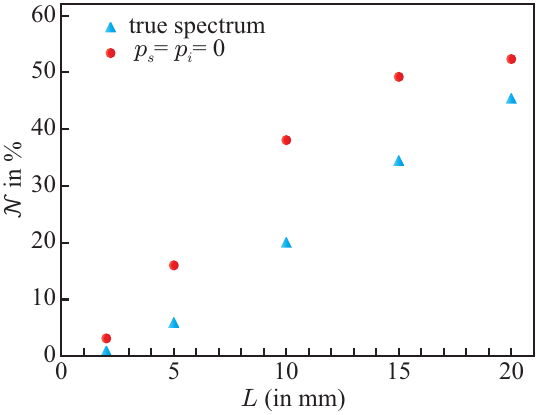}
\caption{ {\bf Dependence of $\mathcal{N}$ on $L$ for two different detection techniques.} The solid circles represent the OAM spectrum with only the $p=0$ mode contribution, while the solid triangles represent the OAM spectrum with OAM contributions summed over all $p$ modes. }
\label{fig:non-conservation-single-p-mode}
\end{figure}

Our present technique works by summing over all the $p$ modes. However, most experiments, testing or harnessing OAM conservation, have employed the OAM detection technique of Ref.~\cite{mair2001nature}, which works by detecting only the $p=0$ modes. So, it becomes natural to ask if the OAM conservation remains intact if only $p=0$ mode contribution is taken. To answer this, we numerically calculate $\mathcal{N}$ as a function of $L$ for two cases: one with only $p=0$ mode contribution and the second with contribution over all the $p$ modes incorporated. The numerical results of Fig.~\ref{fig:non-conservation-single-p-mode} show non-conservation even with only $p=0$ mode contributions. This implies that even with respect to $p=0$ mode detection, the Schmidt-decomposed form given in Eq.~(\ref{two-photon state-Schmidt}) is not correct; it is only a good approximation for very thin crystals.

%Second, experimental configurations typically employed thin crystals in near-collinear geometries where $\alpha_p L \approx 0$. For example, Zeilinger's original experiments used $L=2$ mm crystals, yielding $\alpha_p L$ values small enough that the rotational symmetry-breaking effect was negligible. Thus, the assumption of OAM conservation created an incomplete picture that became foundational to the field despite the absence of rigorous theoretical justification.

%
%
\begin{figure}[hbtp]
\centering
\includegraphics[scale=0.94]{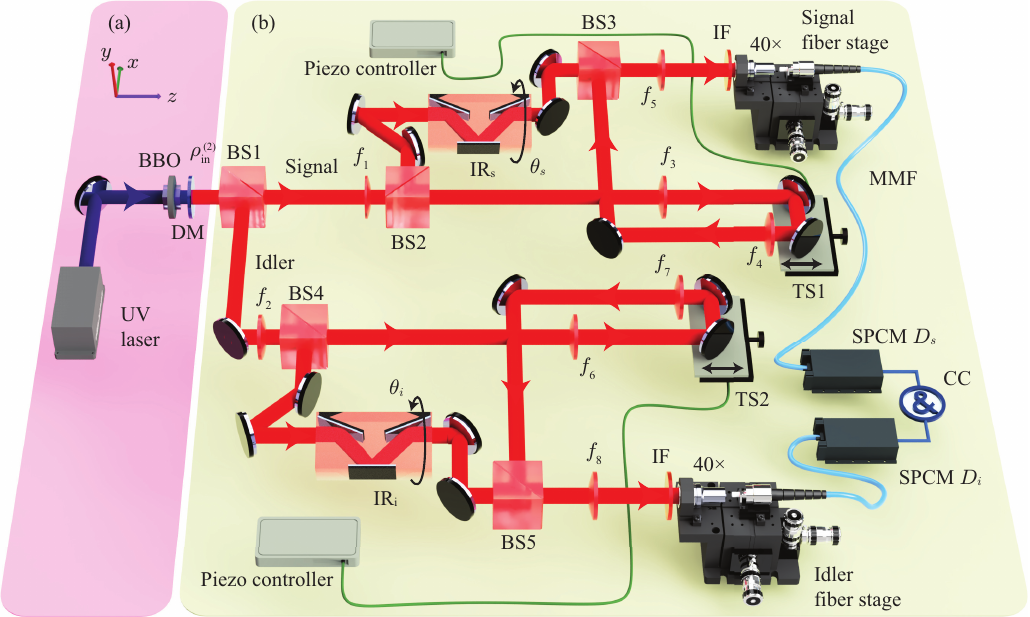}
\caption{{\bf Experimental setup.} (a) Schematic of the experimental setup for generating two-photon state $\rho^{(2)}_{\rm in}$ in the OAM basis. (b) Schematic of the experimental setup of the proposed two-photon OAM spectrum detector for measuring $P^{l_s}_{l_i}$ of $\rho^{(2)}_{\rm in}$. BBO, $\beta$-barium borate crystal; DM, dichroic mirror to block the UV pump and pass the down-converted photon pairs; BS1-BS5, non-polarizing 50:50 beam splitters; ${\rm IR}_s$ and ${\rm IR}_i$, image rotators; TS1 and TS2, translation stages with automated actuators and piezoelectric actuator for finer translation; piezo controller, Arduino UNO board to control the movement of piezo-actuators; IF, interference filter at central frequency 810 nm with FWHM 10 nm; MMF, multimode fiber; SPCM $D_s$ and $D_i$, single photon counting modules for signal and idler; CC, coincidence counting module. }
\label{fig:experimental_setup}
\end{figure}

\noindent{\bf Experimental measurement of two-photon OAM spectrum.} In order to experimentally demonstrate non-conservation of OAM in SPDC, one needs to measure the complete two-photon OAM spectrum $P_{l_s}^{l_i} = \sum^{\infty}_{p_s,p_i =0}\langle|C_{l_i,p_i}^{l_s,p_s}|^2\rangle_e$, with a broadband uniform-efficiency OAM detection techniques that does not involve mode-dependent losses or having prior information about the input state. Figure~\ref{fig:experimental_setup} shows the schematic of our proposed detector for measuring $P_{l_s}^{l_i}$. Figure~\ref{fig:experimental_setup}(a) depicts the setup for generating OAM-entangled two-photon state $|\psi_2\rangle$ which enters a Franson-type two-photon interferometer shown in Fig.~\ref{fig:experimental_setup}(b). The interferometer consists of two image rotators ${\rm IR}_s$ and ${\rm IR}_i$ kept at rotation angles $\theta_s$ and $\theta_i$, and rotates the wavefronts passing through it by $2\theta_s$ and $2\theta_i$, respectively. The angular position of the image rotator is taken to be at  $\theta_{s(i)}=0$ when it is in the $xz$ plane. As indicated in the diagram, we take the direction of  propagation to be along $\hat{z}$, and the $\hat{\bm x}$ direction defines the azimuthal angle $\phi_{s(i)} =0$. First of all, we model the propagation of down-converted photons through the interferometer using a two-photon projection operator $\hat{P}^{(2)}$ (for details, see Supplementary Materials section~\ref{sec:si_tp_path_alternatives})
\begin{align} \label{eqn:projection_P}
\hat{P}^{(2)} = \sum^{\infty}_{l''_s , l''_i = -\infty} \sum^{\infty}_{p''_s , p''_i = 0} e^{i \left[\gamma_1 + \phi_{s1} + \phi_{i1} - \omega_{s0}\left( \tau_{s1} + \tau_{i1}\right) \right]} & \left[|k_1|\hat{\bm x}_s \otimes \hat{\bm x}_i  + |k_2|e^{i\left( \delta + l_s 2\theta_s + l_i 2\theta_i\right)}  \hat{\bm \epsilon}_s\left(\theta_s\right) \otimes \hat{\bm \epsilon}_i\left(\theta_i\right) \right] \nonumber \\
&\times |l''_s, p''_s\rangle_s |l''_i, p''_i\rangle_i {}_s \langle l''_s, p''_s| {}_i \langle l''_i, p''_i|.
\end{align}
Here $k_1 = |k_1|e^{i \gamma_1}$ and $k_2 = |k_2| e^{i \gamma_2}$ are the complex amplitude coupling coefficients that account for the transmission and reflection properties of the beam splitters in each two-photon path alternative. $\delta = \left(\gamma_2 - \gamma_1 \right) + \phi_{s2} + \phi_{i2} - \phi_{s1} - \phi_{i1} - \omega_{s0} \left(\tau_{s2} + \tau_{i2} - \tau_{s1} -\tau_{i1} \right)$, where $\tau_{s1}$, $\tau_{i1}$, $\tau_{s2}$, $\tau_{i2}$ are the photon travel times, $\phi_{s1}, \phi_{i1}, \phi_{s2}, \phi_{i2}$ are the non-dynamical phases, and $\omega_{s0} $ is the frequency of the non-degenerate signal and idler photons. For horizontally polarized signal and idler denoted by $\hat{x_s}$ and $\hat{x_i}$, respectively, the image rotators ${\rm IR}_s$ and ${\rm IR}_i$ change their polarization states to $\hat{\bm \epsilon}_s\left(\theta_s\right)$ and $\hat{\bm \epsilon}_i\left(\theta_i\right)$, respectively, where $\hat{\bm \epsilon}_{s(i)}\left(\theta_{s(i)}\right) = \cos \Psi\left[\theta_{s(i)}\right]\hat{\bm x}_{s(i)} + \sin \Psi\left[\theta_{s(i)}\right]\hat{y}_{s(i)} $. The state $\rho^{(2)}_{\rm out}$ of the down-converted photons exiting the interferometer can thus be written as $\rho^{(2)}_{\rm out} = \hat{P}^{(2)} \rho^{(2)}_{\rm in} \hat{P}^{(2) \dag}$, where $\rho^{(2)}_{\rm in} \equiv \langle |\psi_2\rangle \langle \psi_2|\rangle_e$ is the density matrix of the state entering the interferometer. Using the orthogonality relations $\int^{\infty}_{0}\int^{2\pi}_{0} \left[LG^{|l'_s|}_{p'_s}\left(\rho_s\right) e^{-i l'_s \phi_s}\right]^{*}\left[LG^{|l_s|}_{p_s}\left(\rho_s\right) e^{-i l_s \phi_s}\right] \rho_s d\rho_s d \phi_s = \delta_{l'_s, l_s} \delta_{p'_s,p_s}$ etc. and  the polarization properties $\hat{{\bm x}}_j. \hat{{\bm x}}_j = \hat{{\bm y}}_j. \hat{{\bm y}}_j =1 $ and $\hat{{\bm x}}_j. \hat{{\bm y}}_j =0$ (with $j= s,i$),  we can express the coincidence detection probability $ R^{\delta}_{si} \left(\theta_s, \theta_i\right) = \iint^{\infty}_{0}  \iint^{2 \pi}_{0} {}_s \langle \rho_s, \phi_s| {}_i \langle \rho_i, \phi_i|\rho^{(2)}_{\rm out} |\rho_s, \phi_s \rangle_s |\rho_i, \phi_i \rangle_i ~\rho_s \rho_i d\rho_s d\rho_i d\phi_s d\phi_i $.  at rotation angles $\theta_s$ and $\theta_i$ as (for details, see Supplementary Materials sections~\ref{sec:si_tp_path_alternatives} and \ref{sec:si_coincidence_visibility}):
\begin{align}\label{eqn:R_si_final}
R^{\delta}_{si} \left(\theta_s, \theta_i\right)  =   |k_1|^2 + |k_2|^2 
 +  2|k_1||k_2| \cos\left[ \psi_s\left(\theta_s\right)\right] \cos\left[ \psi_i\left(\theta_i\right)\right]  \sum^{N}_{l_s, l_i =-N}  P^{l_s}_{l_i} \cos\left(\delta + 2l_s \theta_s + 2 l_i \theta_i\right).
\end{align}
In an experimental situation, we encounter two noise sources that affect the coincidence probability. One is the fluctuation in SPDC flux over time because of the intensity fluctuations of pump beam. The second is the background and accidental coincidence counts. We incorporate both of these noise sources, and denote the noise added coincidence detection probability as $ \bar{R}^{\delta}_{si} \left(\theta_s, \theta_i\right)$. We then find a detection scheme for extracting $P^{l_s}_{l_i}$ (see Supplementary Materials section~\ref{sec:si_coincidence_visibility} for details) by measuring $\bar{R}^{\delta}_{si} \left(\theta_s, \theta_i\right)$  at two different settings, $\delta =\delta_c =0$ and $\delta = \delta_d = \pi$. We define the coincidence visibility  as: $\bar{V}_{si}\left(\theta_s, \theta_i\right) = \dfrac{\bar{R}^{\delta=0}_{si} \left(\theta_s, \theta_i\right) - \bar{R}^{\delta= \pi}_{si} \left(\theta_s, \theta_i\right)}{\bar{R}^{\delta=0}_{si} \left(\theta_s, \theta_i\right) + \bar{R}^{\delta= \pi}_{si} \left(\theta_s, \theta_i\right)}$, which can shown to be
\begin{align} \label{eqn:v_si}
\bar{V}_{si}\left(\theta_s, \theta_i\right)=\dfrac{ 2|k_1||k_2| \cos\left[ \psi_s\left(\theta_s\right)\right] \cos\left[ \psi_i\left(\theta_i\right)\right]}{|k_1|^2 + |k_2|^2}  \sum^{N}_{l_s, l_i =-N}  P^{l_s}_{l_i}  \cos\left( 2l_s \theta_s + 2 l_i \theta_i\right). 
\end{align}
We note that the term $\cos\left[ \psi_s\left(\theta_s\right)\right] \cos\left[ \psi_i\left(\theta_i\right)\right]$ in the numerator is because of the fact that image rotators ${\rm IR}_s$ and ${\rm IR}_i$ change the state of polarization of the field passing through it. In order to bypass this effect, we experimentally measure this term and define the polarization-corrected coincidence visibility as:  $V_{si}\left(\theta_s, \theta_i\right) = \bar{V}_{si}\left(\theta_s, \theta_i\right)/ \left[ \cos\left[ \psi_s\left(\theta_s\right)\right] \cos\left[ \psi_i\left(\theta_i\right)\right] \right]$ (see Methods section for details). For the symmetric OAM spectrum, that is,  $P^{l_s}_{l_i} = P^{-l_s}_{-l_i}$, we obtain $\bar{P}^{l_s}_{l_i}$ from $V_{si}\left(\theta_s, \theta_i\right)$ via a two-dimensional Fourier transform as (see detailed calculation in Supplementary Materials section~\ref{sec:si_coincidence_visibility}, also see section~\ref{sec:si_non_symmetric_spectrum_calculation} for the case of asymmetric spectrum):
\begin{equation} \label{eqn:p_ls_li_calculate}
\bar{P}^{l_s}_{l_i} \equiv \iint^{\pi/2}_{-\pi/2} V_{si}\left(\theta_s, \theta_i\right)  \cos\left( 2l_s \theta_s + 2 l_i \theta_i\right) d\theta_s d\theta_i  = \dfrac{2 |k_1||k_2|  }{ |k_1|^2 + |k_2|^2 } P^{l_s}_{l_i}.
\end{equation}
We note that input state is normalized, that is, ${\rm Tr}\left(\rho^{(2)}_{\rm in}\right) =1$. Thus the normalized two-photon OAM spectrum is obtained as $P^{l_s}_{l_i}= \bar{P}^{l_s}_{l_i}/ \left(\sum^{N}_{l_s,l_i = -N} \bar{P}^{l_s}_{l_i}\right)$.  In order to quantify the measurement accuracy of our proposed technique, we use the coefficient of determination as $R^2 = 
\dfrac{\sum^{N}_{l_s,l_i =-N}\left({P_{\rm ob}}^{l_s}_{l_i} - {P_{\rm in}}^{l_s}_{l_i}\right)^2}{\sum^{N}_{l_s,l_i =-N}\left({P_{\rm in}}^{l_s}_{l_i} - \langle {P_{\rm in}}^{l_s}_{l_i}\rangle\right)^2} \times 100\% $, where ${P_{\rm ob}}^{l_s}_{l_i}$ and ${P_{\rm in}}^{l_s}_{l_i}$ are the observed and  input two-photon OAM spectrum, respectively, and $\langle {P_{\rm in}}^{l_s}_{l_i}\rangle$ is the average of ${P_{\rm in}}^{l_s}_{l_i}$.
We note that our measurement technique requires only angle-averaged coincidence detection using two single-pixel SPCMs. It eliminates the need for high spatial-resolution detector, which could be a severe limitation \cite{zia2023natphoton,li2023prl}. Further, unlike the technique of \cite{mair2001nature}, our technique achieves uniform detection efficiency over a broad range of OAM modes and requires no prior knowledge of the input two-photon state.

%We note that this method can also work for non-symmetric $P^{l_s}_{l_i}$, but there we need measurement of coincidence detection probabilities at four interferometric settings $\delta = 0, \pi, \dfrac{\pi}{2}$ and $\dfrac{3\pi}{2}$ instead of just at $\delta = 0$, and $\pi$ (see supplementary material for the detailed calculation of the measurement of non-symmetric $P^{l_s}_{l_i}$ from coincidence measurements at four interferometric settings). 

\section*{Experimental setup and results}
\begin{figure}[htbp]
\centering
\includegraphics[scale=0.87]{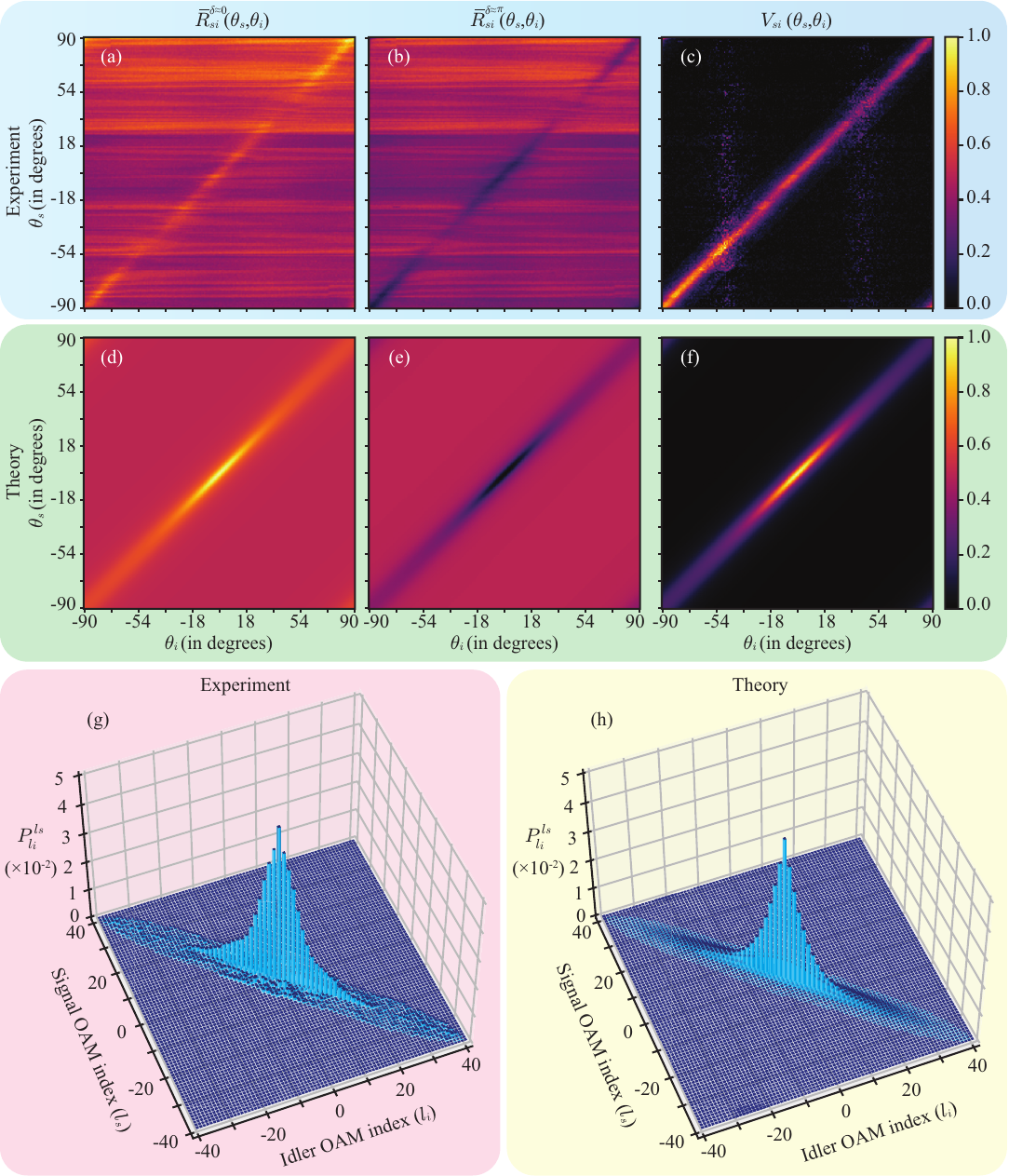}
\caption{{\bf Experimentally measured two-photon OAM spectrum for Type-I SPDC}. (a) Measured $\bar{R}^{\delta}_{si} \left(\theta_s, \theta_i\right)$ at $\delta= \delta_c \approx 0$, and (b) $\delta= \delta_d \approx \pi$. (c) Plot of polarization-corrected coincidence visibility $V_{si}\left(\theta_s, \theta_i\right)$. All are presented as functions of $\theta_s$ and $\theta_i$. (d), (e), and (f) show the corresponding theory plots. (g) Reconstructed $P^{l_s}_{l_i}$ from experimental observations and (h) numerically evaluated $P^{l_s}_{l_i}$ based on Eq.~\ref{eqn:spdc_p_ls_li}.}
\label{fig:result_reconstructed_spectrum}
\end{figure}
Figure~\ref{fig:experimental_setup} (a) presents the schematic of the experimental setup for generating the two-photon state $\rho^{(2)}_{\rm in}$. We employ a $100$ mW TOPTICA Topmode continuous-wave UV pump laser at $405$ nm wavelength, attenuated to $5$ mW using neutral density filters. The pump beam has a beam waist of $388~\mu$m and the coherence length of $25$ meters. It passes through a BBO crystal with thickness $L = 15$ mm mounted on a goniometer, with $\theta_p = 28.65^{\circ}$ to achieve Type-I collinear phase-matching. A dichroic mirror is placed immediately after the nonlinear crystal to block the UV pump while transmitting the down-converted photon pairs. The down-converted two-photon state $\rho^{(2)}_{\rm in}$ enters the interferometer shown in Fig.~\ref{fig:experimental_setup}(b) and passes through a non-polarizing $50:50$ beam splitter (BS1). The signal and idler photons go through two separated Mach-Zehnder interferometers with image rotators oriented at $\theta_s$ and $\theta_i$, respectively. The signal and idler photons pass through interference filters of central wavelength $810$ nm and FWHM $10$ nm and are collected using multimode fibers (MMF) of core diameters $62 ~\mu$m and Newport $40\times$ objective lens place on XYZ fiber stages. For the signal interferometer arm containing $IR_s$, lenses $f_1=400$ mm and $f_5=500$ mm along with the objective lenses, image the Fourier transform of the crystal plane on to the tip of the MMF with some demagnification. For the other arm, which is 80 cm longer, lenses  $f_3=200$ mm and $f_4 = 200$ mm kept in a $4f$ configuration do the imaging with the same magnification. Similar detection arrangement is made in the idler interferometer. The photons are then detected in coincidence using two Excelitas single photon counting modules (SPDMs), $D_s$ and $D_i$.  We employ a PicoQuant TimeHarp 260 (PICO model) for coincidence detection. In the above interferometer there are four alternative pathways by which the signal and idler photons can reach the two detectors, $D_s$ and $D_i$. In order to eliminate two of the alternative pathways, we keep the arms containing the image rotator to be $80$ cm shorter than the other arm of the interferometers, and choose the coincidence time-window appropriately (see Methods section ). Additionally, we ensure two-photon interference by matching the path lengths (see Methods for experimental details).

Figure~\ref{fig:result_reconstructed_spectrum} presents our experimental results. Fig.~\ref{fig:result_reconstructed_spectrum}(a) shows the measured coincidence detection probability  $\bar{R}^{\delta}_{si} \left(\theta_s, \theta_i\right)$ as a function of $\theta_s$ and $\theta_i$  for $\delta  = \delta_c \approx 0$. The corresponding plot for  $\delta  = \delta_d \approx \pi$ is shown in Fig.~\ref{fig:result_reconstructed_spectrum}(b) (see Methods for measurement details). %For both plots, coincidence counts per second are normalized to the maximum count rate of $\bar{R}^{\delta=\delta_c}_{si} \left(\theta_s, \theta_i\right)$ ????????. 
The polarization-corrected coincidence visibility $V_{si}\left(\theta_s, \theta_i\right)$  is plotted in Fig.~\ref{fig:result_reconstructed_spectrum}(c), with angular resolution and sampling requirements discussed in Methods. Figures~\ref{fig:result_reconstructed_spectrum}(d), \ref{fig:result_reconstructed_spectrum}(e), \ref{fig:result_reconstructed_spectrum}(f) show the corresponding theory plots, which account for experimental collection aperture effects due to coupling into multimode fibers (see Supplementary Materials section~\ref{sec:si_effect_of_clipping}). The reconstructed $(81\times 81)$-dimensional $P^{l_s}_{l_i}$ using our technique is displayed as a sky-blue bar plot in Fig.~\ref{fig:result_reconstructed_spectrum}(g), with the corresponding theory plot shown in Fig.~\ref{fig:result_reconstructed_spectrum}(h). The measurement accuracy is found to be $R^2 = 87.18\%$, demonstrating good agreement between theory and experiments. We find significant off-diagonal elements in our measurement of $P^{l_s}_{l_i}$, directly demonstrating OAM non-conservation in type-I SPDC. The experimentally measured $\mathcal{N} = 42.92\%$ matches closely with the theoretically predicted value of $\mathcal{N} = 34.40\%$.

\section*{Discussion}

In summary, we have proposed and experimentally demonstrated a high-sensitivity detector for measuring the joint OAM spectrum of entangled two-photon fields produced by spontaneous parametric down-conversion. Our detector measures the two-photon OAM spectrum with uniform efficiency and without prior knowledge of the input state. The high-fidelity two-photon OAM detector demonstrated in our work not only removes one of the roadblock for realizing high-dimensional quantum advantages  but has also proved central to resolving a foundational question, namely, the conservation of OAM in the SPDC process. The current understanding regarding OAM conservation is that although it is not conserved in Type-II SPDC and even in extremely non-collinear Type-I SPDC, it remains conserved in nearly-collinear Type-I SPDC. This understanding is based on experimental results obtained with limited-sensitivity OAM detectors and theoretical models that employ severe phase-matching approximations. Using our detector, contrary to the current understanding, we have experimentally demonstrated non-conservation of OAM even in Type-I SPDC, with the non-conservation parameter $\mathcal{N} = 42.92\%$. We have presented a rigorous theoretical framework free of standard phase-matching approximations and have shown that our experimental observations match with the existing theory with up to 87$\%$ fidelity. Furthermore, we have presented a detailed analysis showing that the transverse walk-off due to the anisotropy on the nonlinear crystal is at the root of OAM non-conservation. The reason why the past theory work could not predict non-conservation is because of using the standard phase-matching approximations. Conservation of OAM in Type-I SPDC has been the underlying assumption in several widely-used techniques for generating OAM-based maximally-entangled states for high-dimensional quantum key distribution, entanglement-based quantum computing, and fundamental tests \cite{karan2023prapplied, vaziri2003prl, wang2017optica, liu2020sciadvances, qiu2023natcomm, zhang2017natcomm}. Therefore, we expect our work to have important implications not only for the detection of OAM-entangled states but also for more accurate generation of custom OAM-entangled states for quantum technologies.

%These findings have immediate implications for quantum technologies employing type-I SPDC for the generation of entangled photon pairs in the OAM basis. Protocols assuming perfect OAM conservation—including high-dimensional quantum key distribution, entanglement-based quantum computing, and fundamental tests--require reanalysis. However, this also enables optimization: crystal parameters can be tuned via $\alpha_p L$ to control the degree of OAM conservation or non-conservation for specific applications.   

%While restricted to diagonal elements (OAM-spectrum) rather than full density matrix reconstruction, this capability enables accurate characterization of OAM correlations in SPDC-generated states. Combined with complementary tomographic techniques, this detector could provide a foundation for complete high-dimensional entangled state characterization, opening new avenues for testing fundamental quantum mechanics and developing OAM-based quantum technologies.

%Beyond SPDC, this understanding can be extended to any nonlinear process in anisotropic media where pump walk-off occurs, including sum-frequency generation, four-wave mixing, and optical parametric amplification, suggesting similar symmetry-breaking effects in angular momentum transfer across nonlinear optics.

%%%%%%%%%%%%%%%%%%%%%%%Methods%%%%%%%%%%%%%%%%%%%
\section*{Materials and Methods}\label{sec:methods}
\noindent{\bf Calibration of the polarization effect of the interferometer}\label{sec:si_pol_calibration}\\
To calibrate the polarization changes induced in the output of the interferometer shown in Fig.~\ref{fig:experimental_setup}(b) due to the rotation of ${\rm IR}_s$ and ${\rm IR}_i$, one needs to measure $\dfrac{1}{ \left[ \cos\left[ \psi_s\left(\theta_s\right)\right] \cos\left[ \psi_i\left(\theta_i\right)\right] \right]}$. Here, this measurement is conducted in two stages. First, we obtain $\dfrac{1}{\cos\left[ \psi_s\left(\theta_s\right)\right]} = \sqrt{\dfrac{I_{s2}^{\rm tot}}{I_{s2x}(\theta_s)}}$ by measuring $I_{s2}^{\rm tot}$ and $I_{s2x}(\theta_s)$. To do this, we block all arms of the interferometer except the one containing ${\rm IR}_s$, allowing light to pass only through ${\rm IR}_s$. We replace the signal fiber stage with an EMCCD camera to measure the transmitted state from ${\rm IR}_s$. The total intensity $I_{s2}^{\rm tot}$, which is independent of $\theta_s$, is obtained by summing over all EMCCD camera pixel values. To measure $I_{s2x}\left(\theta_s\right)$, we place a polarizer with its axis along the $\hat{\bm x}$ direction just after ${\rm IR}_s$ to perform projective measurements along $\hat{\bm x}_s$. We then measure the sum of all EMCCD camera pixel values at various rotation angles $\theta_s$ of ${\rm IR}_s$.

Next, we obtain $ \dfrac{1}{  \cos\left[ \psi_i\left(\theta_i\right)\right] } = \sqrt{\dfrac{I_{i2}^{\rm tot}}{I_{i2x}(\theta_i)}}$ by measuring $I_{i2}^{\rm tot}$ and $I_{i2x}\left(\theta_i\right)$. The process is similar to the previous case, but we now block all interferometer arms except the one containing ${\rm IR}_i$. We place the EMCCD camera at the idler fiber stage and measure $I_{i2}^{\rm tot}$ by summing over all pixel values. Subsequently, we place a polarizer just after ${\rm IR}_i$ with its axis along $\hat{\bm x}$ to project along $\hat{\bm x}_i$. We then measure the intensity using the EMCCD camera at various rotation angles $\theta_i$ of ${\rm IR}_i$.
\begin{figure}[!h]
\centering
\includegraphics[scale=0.95]{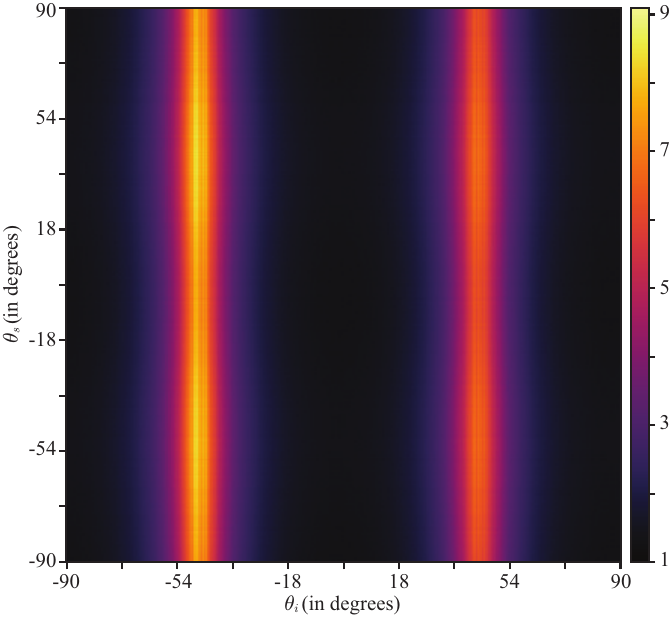}
\caption{{\bf Plot of polarization calibration factor as a function of ${\rm IR}_s$ and ${\rm IR}_i$ angles.} The calibration factor $\dfrac{1}{\cos[\psi_s(\theta_s)] \cos[\psi_i(\theta_i)]}$ is plotted as a function of rotation angles $\theta_s$ and $\theta_i$ of ${\rm IR}_s$ and ${\rm IR}_i$, respectively.}
\label{fig:si_polarization_calibration_2d}
\end{figure}
For all measurements, we subtract background noise by taking EMCCD readings in the absence of any input light. This process yields two one-dimensional arrays that represent $\dfrac{1}{  \cos\left[ \psi_s\left(\theta_s\right)\right] }$ and $ \dfrac{1}{  \cos\left[ \psi_i\left(\theta_i\right)\right] }$ as functions of $\theta_s$ and $\theta_i$, respectively. We then combine these results to obtain the two-dimensional function  $ \dfrac{1}{ \left[ \cos\left[ \psi_s\left(\theta_s\right)\right] \cos\left[ \psi_i\left(\theta_i\right)\right] \right]}$ by taking the outer product of the two one-dimensional arrays. 
Figure~\ref{fig:si_polarization_calibration_2d} presents this measured function for our setup, with $\theta_s$ and $\theta_i$ ranging from $-90^\circ$ to $90^\circ$, which is our region of interest. We use this calibration result to calculate $V_{si}\left(\theta_s, \theta_i\right)$ and obtain the results shown in Fig.~\ref{fig:result_reconstructed_spectrum}(c) using  Eq.~(\ref{eqn:p_ls_li_calculate}).
\\
\\
\\
\noindent{\bf Selection of proper coincidence time-window}\label{sec:si_coin_time_window}\\
\begin{figure}[!b]
\centering
\includegraphics[scale=0.85]{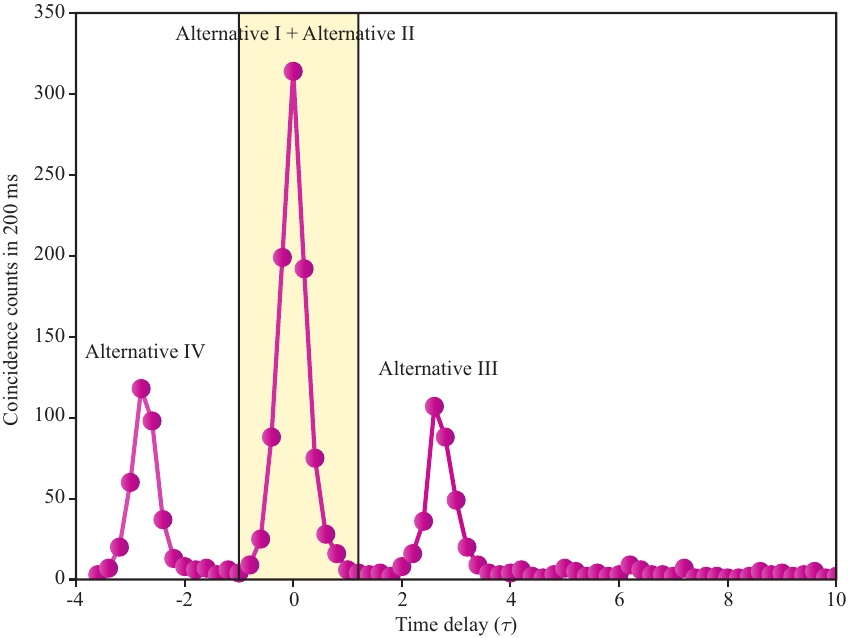}
\caption{{\bf Coincidence count as a function of time delay between signal and idler detectors.} Plot of coincidence count collected over 200 ms versus time delay $\tau$ between detectors $D_s$ and $D_i$, respectively. The yellow shaded region (width = 2 ns) corresponds to the temporal overlap between two-photon path alternatives I and II of the interferometer presented in Fig.~\ref{fig:experimental_setup}(b).}
\label{fig:si_coincidence_window}
\end{figure}
In Fig.~\ref{fig:experimental_setup}(b), if detector $D_s$ registers an event at time $t$ and detector $D_i$ registers an event at time $t+\tau$, then this can be considered a coincidence event at instant $t$ within the given coincidence window $\tau$. The number of such events is called the coincidence count at that instant. Figure~\ref{fig:si_coincidence_window} shows the plot of the coincidence count as a function of $\tau$ in our experiment when $\theta_s =0$ and $\theta_i =0$. Coincidence counts are obtained for a collection time of $200$ ms. We observe three peaks, each with a linewidth determined primarily by the timing resolution of the SPCM. The central peak results from the contributions of Alternatives I and II, where both signal and idler arrive at $D_s$ and $D_i$, respectively, at the same time. The rightmost peak corresponds to Alternative III, as the idler takes longer to reach $D_i$ than the signal takes to reach $D_s$. In our setup, the idler path is 80 cm longer than the signal path. Similarly, the leftmost peak corresponds to Alternative IV, where the signal path is 80 cm longer than the idler path. Here, we define $\tau =0$ with respect to the central peak, with negative $\tau$ values indicating that the signal arrives earlier than the idler. The distances of the side peaks from the central peak can be tuned by adjusting the path length difference between the longer and shorter paths. For our experiment, 80 cm is sufficient to clearly distinguish the side peaks from the central peak, with a bin width of 200 picoseconds.

Since our theoretical framework for two-photon interference considers only the contributions of Alternatives I and II, we select a $\tau$ range from $-1$ ns to $1$ ns, shown as the yellow shaded region in Fig.~\ref{fig:si_coincidence_window}. The total coincidence count is defined as the sum of counts within this range, thereby eliminating the contributions of Alternatives III and IV through the use of a 2 ns coincidence time-window. We note that the total coincidence count includes some accidental counts. To subtract these experimentally, we measure the coincidence count at sufficiently large $\tau$ values. For instance, in Fig.~\ref{fig:si_coincidence_window}, coincidence counts at  $\tau > 8$ ns are non-zero, representing accidental coincidences. We average these values at $\tau \geq 8$ ns and subtract this average from each point in the plot. The sum of counts within the yellow-shaded region after this accidental coincidence count subtraction represents the real coincidence count.
%Finally, with this coincidence count configuration, we obtain the experimental results shown in Fig.~2.
\\
\\
\noindent{\bf Matching path length for two-photon interference}\label{sec: si_matchpathlength}\\
\begin{figure}[!t]
\centering
\includegraphics[scale=0.8]{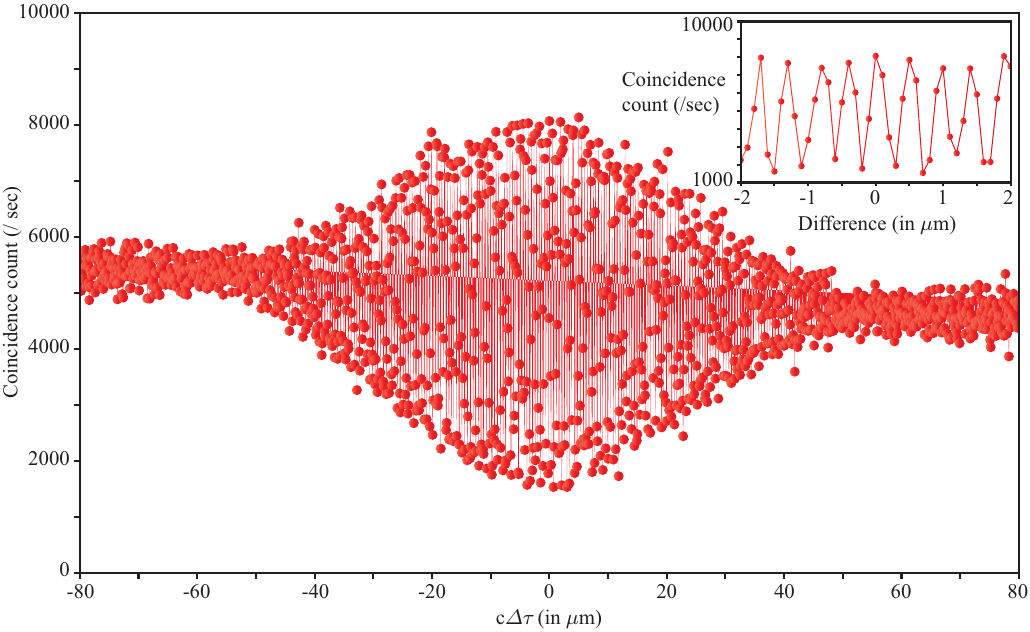}
\caption{{\bf Coincidence count rates versus two-photon path length difference.} Measured coincidence count rates $\bar{R}_{si}(\theta_s, \theta_i)$ as a function of two-photon path length difference $c\Delta\tau$ when $\theta_s = \theta_i = 0^{\circ}$. The inset shows the same data over a narrower range ($-2$ to $+2~\mu$m) to highlight the oscillatory behavior of $\bar{R}_{si}(\theta_s, \theta_i)$.}
\label{fig:si_hom_temporal_plot}
\end{figure}
Two-photon interference requires coherence property to be satisfied in all degrees of freedom. If the degree of coherence is zero in any one degree of freedom, interference will not occur. In our setup, we explore two-photon angular coherence. Therefore, we must ensure that coherence properties for polarization and temporal degrees of freedom are satisfied. The coherence property of the two-photon state in the polarization degree of freedom is inherently satisfied here as the pump photon is perfectly polarized \cite{kulkarni2016pra}. To ensure coherence in the spatial degree of freedom, we align  both Mach-Zehnder-type interferometers to  the zero fringe condition individually  and set $\theta_s =0$ and $\theta_i =0$. Next, we need to examine the temporal degree of freedom. Following the formalism presented in Ref.~\cite{jha2008pra}, two-photon temporal interference depends on two parameters, $\Delta \tau$ and $\Delta \tau'$. From the schematic of the path alternatives shown in Fig.~\ref{fig:si_path_alternatives} in the Supplementary Materials section~\ref{sec:si_tp_path_alternatives}, these parameters can be expressed as:
\begin{equation}
\Delta \tau = \left(\tau_p + \dfrac{\tau_{s1}+ \tau_{i1}}{2} \right) - \left(\tau_p + \dfrac{\tau_{s2}+ \tau_{i2}}{2} \right),
\end{equation}
\begin{equation}
\Delta \tau' =\left( \tau_{s1} - \tau_{i1}\right) - \left(\tau_{s2}- \tau_{i2}\right).
\end{equation}
The condition for temporal coherence is that $\Delta \tau < \tau^{p}_{\rm coh}$ and $\Delta \tau'   < \tau^{(2)}_{\rm coh}$, where $\tau^{p}_{\rm coh}$ is the temporal coherence time of the pump and $\tau^{(2)}_{\rm coh}$  is the two-photon temporal coherence time. For our experimental setup, the pump coherence length is 
$c \tau^{p}_{\rm coh} \approx 25$ m, where $c$ is the velocity of light in free space, and the two-photon coherence length $c \tau^{(2)}_{\rm coh} $ is approximately $60~\mu$m. In order to satisfy these conditions for two-photon temporal interference, we scan the interferometer by changing the path length of the longer arm of the upper interferometer. This is accomplished using TS1, an automated translation stage with a step size of $0.2~\mu$m, which changes $\tau_{s1}$ and consequently modifies both $\Delta \tau$  and $\Delta \tau'$.

Figure~\ref{fig:si_hom_temporal_plot} shows the plot of the coincidence count per second as a function of the two-photon path length difference $c \Delta \tau$. The plot reveals that the coincidence count follows an exponentially decaying oscillatory function, which is the signature of the temporal Hong-Ou-Mandel interference pattern. The inset of Fig.~\ref{fig:si_hom_temporal_plot} highlights the oscillatory nature, which is plotted over a small range of $c \Delta \tau$. Finally, for our measurements of coincidence visibility as functions of $\theta_s$ and $\theta_i$, we fix TS1 at the position where $c \Delta \tau = 0$, thereby ensuring maximum two-photon temporal coherence.
\\
\\
\noindent{\bf Measurement of coincidence visibility}\label{sec:si_coin_visibility_measurement}\\
In order to calculate the coincidence visibility $\bar{V}_{si}\left(\theta_s, \theta_i\right)$, we need to measure the coincidence detection probability $\bar{R}^{\delta}_{si} \left(\theta_s, \theta_i\right)$ at $\delta =0$ and $\pi$ for every $\theta_s$ and $\theta_i$. Usually, this requires keeping $\delta$ stable throughout the measurement, which is challenging and needs special active stabilization techniques. However, we use a simpler method to bypass this issue. We find that, for probability distributions $P^{l_s}_{l_i}$ generated from SPDC, $\bar{R}^{\delta}_{si}\left(\theta_s, \theta_i\right)$ has a useful property: it reaches its maximum at $\delta =0$ and minimum at $\delta = \pi$, for each $\theta_s$ and $\theta_i$. Using this fact, we define $  \left\lbrace \bar{R}^{\delta}_{si}\left(\theta_s, \theta_i\right) \right \rbrace_{\rm max }= \bar{R}^{\delta =0}_{si}\left(\theta_s, \theta_i\right) \equiv  \bar{R}^{\rm max}_{si}\left(\theta_s, \theta_i\right)$  and   $  \left\lbrace \bar{R}^{\delta}_{si}\left(\theta_s, \theta_i\right) \right \rbrace_{\rm min }= \bar{R}^{\delta =\pi}_{si}\left(\theta_s, \theta_i\right) \equiv  \bar{R}^{\rm min}_{si}\left(\theta_s, \theta_i\right)$ for all $\theta_s$ and $\theta_i$.

This way, we do not need to track the phase exactly. We need only determine $\bar{R}^{\rm max}_{si}\left(\theta_s, \theta_i\right)$ and $\bar{R}^{\rm min}_{si}\left(\theta_s, \theta_i\right)$ for each $\theta_s$ and $\theta_i$. To do this, we measure $\bar{R}^{\delta}_{si}\left(\theta_s, \theta_i\right)$ at $40$ different $\delta$ values, in steps of $9^{\circ}$. This covers the full $2\pi$ range for $\delta$, ensuring we capture both $\delta =0$ and $\delta =\pi$, regardless of the initial $\delta$ value. From these $40$ measurements, we can find the maximum and minimum to calculate $\bar{V}_{si}\left(\theta_s, \theta_i\right)$. We use the translation stage TS1 (see Fig.~\ref{fig:experimental_setup}(b)) with a piezo actuator embedded  with it to change $\delta$. Moving the stage by $202.5$ nm changes the phase by $2\pi$. This is because the movement of the translation stage by a distance $x$ changes the path length by $2x$, and we need a path length change of $405$ nm, which is the wavelength of our pump laser for a $2\pi$ phase shift.

We achieve this precise displacement using a Thorlabs PC4QRCo-Fired piezo actuator mounted on TS1. We calibrate the piezo and determine that a voltage change of $117.6$ mV to the piezo corresponds to a $\delta$ change of $9^{\circ}$. We use an Arduino Uno Rev3 to control the piezo actuator, varying the voltage from $0$ to $4.7$ V in $117.6$ mV steps to obtain $40$ measurement values. This setup allows us to accurately measure $\bar{R}^{\delta}_{si}\left(\theta_s, \theta_i\right)$ at each step, enabling us to determine $\bar{R}^{\rm max}_{si}\left(\theta_s, \theta_i\right)$ and $\bar{R}^{\rm min}_{si}\left(\theta_s, \theta_i\right)$ for every $\theta_s$ and $\theta_i$ combination. We repeat this measurement process for $\theta_s$ and $\theta_i$ ranging from $-90^{\circ}$ to $90^{\circ}$. Plots of experimentally measured $\bar{R}^{\delta=0}_{si}\left(\theta_s, \theta_i\right)$ and $\bar{R}^{\delta=\pi}_{si}\left(\theta_s, \theta_i\right)$ are presented in Figs.~\ref{fig:result_reconstructed_spectrum}(a) and \ref{fig:result_reconstructed_spectrum}(b), respectively. These enable us to calculate the coincidence visibility through the following relation: $\bar{V}_{si}\left(\theta_s, \theta_i\right) = \dfrac{\bar{R}^{\delta=0}_{si} \left(\theta_s, \theta_i\right) - \bar{R}^{\delta= \pi}_{si} \left(\theta_s, \theta_i\right)}{\bar{R}^{\delta=0}_{si} \left(\theta_s, \theta_i\right) + \bar{R}^{\delta= \pi}_{si} \left(\theta_s, \theta_i\right)}$. 
\\
\\
\noindent{\bf Angular resolution in two-photon OAM spectrum measurement}\label{sec:si_angular_resol_tp_spectrum}\\
Our technique of measuring $P^{l_s}_{l_i}$ requires a two-dimensional Fourier transform of $V_{si}\left(\theta_s, \theta_i\right)$  as shown in Eq.~(\ref{eqn:p_ls_li_calculate}). This transformation requires appropriate sampling resolution.  According to the Whittaker-Shannon sampling theorem \cite{goodman2005}, the resolution of $\theta_s$ and $\theta_i$ should be smaller than $\frac{180^{\circ}}{2l^{\rm max}_{s} +1}$ and $\frac{180^{\circ}}{2l^{\rm max}_{i} +1}$ respectively, where $l^{\rm max}_{s}$ and $l^{\rm max}_{i}$ are the highest OAM mode indices of signal and idler photons. In our experiment, since $l^{\rm max}_{s} = l^{\rm max}_{i}$, we require $V_{si}\left(\theta_s, \theta_i\right)$ with minimum dimensions of $(2l^{\rm max}_{s} + 1) \times (2l^{\rm max}_{s} + 1)$ to accurately reconstruct $P^{l_s}_{l_i}$.

Since we do not use any prior information about the input state, including its dimensionality, we first measure the coincidence visibility $\bar{V}_{si}\left(\theta_s, \theta_i\right)$ with a larger step size for $\theta_s$ and $\theta_i$ to obtain a rough estimation of the state's dimension. Subsequently, in the actual measurement, we use this information to determine the appropriate step size for $\theta_s$ and $\theta_i$ and measure $\bar{V}_{si}\left(\theta_s, \theta_i\right)$ more precisely. For the results presented in Fig.~\ref{fig:result_reconstructed_spectrum}, we have chosen a step size of $0.9^{\circ}$ for both $\theta_s$ and $\theta_i$.

\bibliography{ref_two_photon_oam_detector} % for a file named science_template.bib
\bibliographystyle{sciencemag}

%%%%%%%%%%%%%%%% ACKNOWLEDGEMENTS %%%%%%%%%%%%%%%

\section*{Acknowledgments} 

We acknowledge fruitful discussions with Prof. Martin P. van Exter from Leiden University, The Netherlands on this work.

\paragraph*{Funding:} We acknowledge financial support from the Science and Engineering Research Board through grants STR/2021/000035 \& CRG/2022/003070, and from the Department of Science \& Technology, Government of India through grant DST/ICPS/QuST/Theme-­1/2019 and through the National Quantum Mission (NQM) technical group project on quantum imaging.

\paragraph*{Author contributions:} S.K. and A.K.J proposed and developed the idea. S.K. performed the experiments with help from A.K.J. S.K. and A. K. J wrote the manuscript. A.K.J. supervised the overall work.

\paragraph*{Competing interests:}
The authors declare that they have no competing interests.

\paragraph*{Data and materials availability:} All data needed to evaluate the
conclusions in the paper are present in the paper and/or in the Supplementary Materials.

%%%%%%%%%%%%%%%% SUPPLEMENT LIST %%%%%%%%%%%%%%%

\subsection*{Supplementary Materials}
{\bf This PDF file includes:}
Supplementary Sections S1 to S5\\
Figs. S1 to S3\\
%%Tables S1 to S4\\
References \textit{(1-\arabic{enumiv})}\\ 
%%%%%%%%%%%%%%%% END OF MAIN TEXT %%%%%%%%%%%%%%%

\newpage

%%%%%%%%%%%%%%%% START OF SUPPLEMENT %%%%%%%%%%%%%%%

% Figures, tables, equations and pages in the supplement are numbered S1, S2 etc.
\renewcommand{\thefigure}{S\arabic{figure}}
\renewcommand{\thetable}{S\arabic{table}}
\renewcommand{\theequation}{S\arabic{equation}}
\renewcommand{\thepage}{S\arabic{page}}
\renewcommand{\thesection}{S\arabic{section}}
\setcounter{figure}{0}
\setcounter{table}{0}
\setcounter{equation}{0}
\setcounter{page}{1} % not 0 as \newpage already started a supplementary page
% References continue the numbering from the main text.

%%%%%%%%%%%%%%%% SUPPLEMENT TITLE PAGE %%%%%%%%%%%%%%%

\begin{center}
\section*{Supplementary Materials for\\ \scititle}

Suman Karan$^{\ast}$,
Anand K. Jha\\ % we're not in a \author{} environment this time, so use \\ for a new line
\small$^\ast$Corresponding author. Email: sumankaran2@gmail.com, akjha@iitk.ac.in
\end{center}

\subsubsection*{This PDF file includes:}
Supplementary Sections S1 to S5\\
Figures S1 to S3\\

%%\subsubsection*{Other Supplementary Materials for this manuscript:}

\newpage
%
%
%
%
%
%
%
%
%
%%%%%%%%%%%% CAPTIONS FOR OTHER SUPPLEMENTARY FILES %%%%%%%%%%
%
\clearpage % Clear all remaining figures and tables then start a new page
%
%
%

%\section*{Supplementary Text}

\section{Calculation of spatial walk-off angle}\label{sec:si_alphap_zetap}
\begin{figure}[hbtp]
\centering
\includegraphics[scale=0.9]{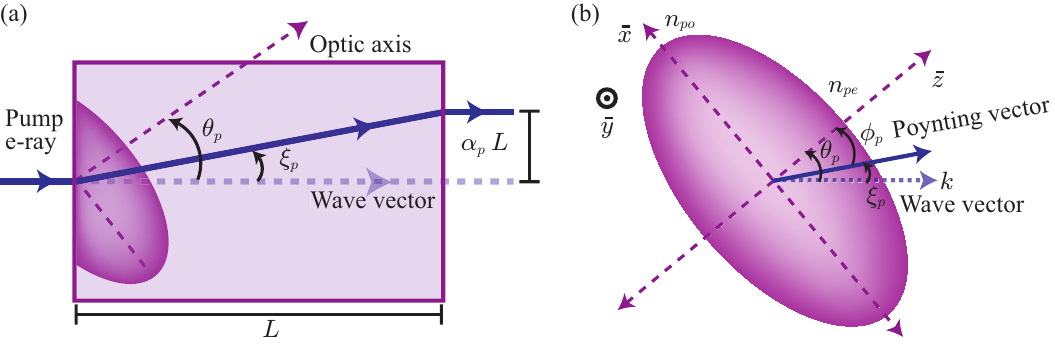}
\caption{{\bf Spatial walk-off in anisotropic crystals.} (a) Schematic of the spatial walk-off experienced by the extraordinary-polarized pump beam propagating through a negative uniaxial BBO crystal. (b) Index ellipse representation in the principal coordinate frame $\bar{x}\bar{y}\bar{z}$. $\theta_p$ is the angle between the crystal optic axis ($\bar{z}$-axis) and the pump wave vector direction. $\phi_p$ is the angle between the $\bar{z}$-axis and the Poynting vector direction. The walk-off angle is $\zeta_p = \phi_p - \theta_p$.}
\label{fig:si_walkoff_index_ellipse}
\end{figure}

The spatial walk-off phenomenon in anisotropic crystals arises from the geometric properties of wave propagation in uniaxial media. In a negative uniaxial crystal like BBO, the crystal optic axis makes an angle $\theta_p$ with the pump propagation direction, as shown in Fig.~\ref{fig:si_walkoff_index_ellipse}(a). For an extraordinary-polarized beam, the wave vector direction ${\bm k}$ and the Poynting vector direction (energy flow) differ by an angle $\zeta_p$, as illustrated in Fig.~\ref{fig:si_walkoff_index_ellipse}(a). In order to calculate $\tan \zeta_p$, we first define the pump wave vector direction as:
\begin{align}\label{eq:si_tantheta}
\tan \theta_p = \dfrac{q_{\bar{x}}}{k_{\bar{z}}},
\end{align}
where $\left( \bar{x}, \bar{y}, \bar{z} \right)$ denotes the principal coordinate frame of the negative uniaxial crystal with the optic axis along $\bar{z}$. Figure~\ref{fig:si_walkoff_index_ellipse}(b) shows these coordinates and the index ellipse geometry explicitly. Since $\theta_p$ lies in the $\bar{x}\bar{z}$ plane and the crystal is uniaxial, we focus on the dispersion relation in this plane:
\begin{align}\label{eq:si_dispersion_rel}
\dfrac{q^2_{\bar{x}}}{n^2_{pe}} + \dfrac{k^2_{\bar{z}}}{n^2_{po}} = \dfrac{\omega_p^2}{c^2}.
\end{align}
Here $n_{po}$ and $n_{pe}$ are the ordinary and extraordinary refractive indices at the pump wavelength, $\omega_p$ is the pump frequency, and $c$ is the speed of light in vacuum.

The normal to the tangent of the k-space ellipse in Eq.~(\ref{eq:si_dispersion_rel}) defines the Poynting vector direction. If the angle between the $\bar{z}$-axis and the Poynting vector is $\phi_p$, then using Eqs.~(\ref{eq:si_tantheta}) and~(\ref{eq:si_dispersion_rel}), $\tan \phi_p$ can be expressed as:
\begin{align}\label{eq:si_tanphip}
\tan{\phi_p} = -\dfrac{d k_{\bar{z}}}{d q_{\bar{x}}} =\dfrac{n^2_{po}}{n^2_{pe}} \tan \theta_p.
\end{align}
Since $\zeta_p = \phi_p - \theta_p$, using Eqs.~(\ref{eq:si_tantheta}) and~(\ref{eq:si_tanphip}), we obtain:
\begin{align}\label{eq:si_express tanzetap}
\tan \zeta_p = \tan \left(\phi_p - \theta_p \right)
= \dfrac{\tan \phi_p - \tan \theta_p}{ 1 +  \tan \phi_p \tan \theta_p} = \frac{(n^2_{po}- n^2_{pe})\sin\theta_p \cos\theta_p}{n^2_{po}\sin^2\theta_p + n^2_{pe} \cos^2 \theta_p},
\end{align}
which is essentially $\alpha_p$ in the phase-matching term shown in Eq.~(\ref{eq: alphap_phasematch}) in the main text. Thus, we derive that $\alpha_p = \tan \zeta_p $.

\section{Two-photon path alternatives and derivation of  interferometer projection operator}\label{sec:si_tp_path_alternatives}
When a two-photon state $\rho^{(2)}_{\rm in}$ passes through the first beam splitter of the orbital angular momentum (OAM) detector shown in Fig.~4(b) of the main text, three scenarios are possible: (i) both photons reflect toward the signal arm, (ii) both photons transmit toward the idler arm, or (iii) one photon reflects while the other transmits, resulting in one photon traveling toward each arm. Coincidence detection is only achievable in the third scenario, since scenarios (i) and (ii) always result in zero counts at one of the single-photon counting modules  (SPCMs). Consequently, our analysis of coincidence detection focuses exclusively on the third scenario.

\begin{figure}[hbtp]
\centering
\includegraphics[scale=1]{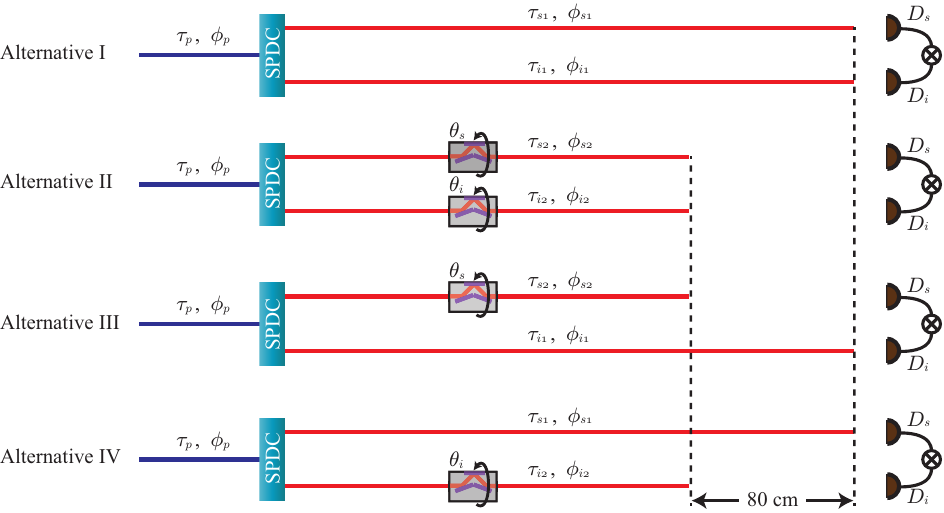}
\caption{{\bf Unfolded two-photon path alternatives for the interferometer setup shown in Fig.~4(b) in the main text.} The signal and idler photons pass through the two-photon interferometer and are detected in coincidence at detectors $D_s$ and $D_i$ via four possible two-photon paths shown schematically.}
\label{fig:si_path_alternatives}
\end{figure}

Under these conditions, there are four possible pathways through which the signal and idler photon pair can reach detectors $D_s$ and $D_i$ as shown in Fig.~4(b).  These are termed two-photon path alternatives for the interferometer and  are illustrated in  Figure~\ref{fig:si_path_alternatives}. Here, $\tau_{s1}$, $\tau_{s2}$  denote signal photon travel times to reach $D_s$ through the longer and shorter paths, respectively, with corresponding non-dynamical phases $\phi_{s1}$ and $\phi_{s2}$. Similarly, for the idler photon, $\tau_{i1}$ and $\tau_{i2}$ represent  the travel times to  reach $D_i$, with associated non-dynamical phases  $\phi_{i1}$  and $\phi_{i2}$. The pump photon follows a single path to the crystal with travel time $\tau_p$ and non-dynamical phase $\phi_p$.

In our experimental design, we set the path length difference between the longer and shorter arms to 80 cm. This requires a sufficiently large coincidence time-window to register coincidence events from alternatives III and IV, where either the signal or idler photon takes a longer time to reach its detector, preventing simultaneous arrival of the photon pair. In contrast, for alternatives I and II, both photons arrive simultaneously at detectors $D_s$ and $D_i$. Exploiting this temporal distinction, we choose a suitably narrow coincidence time-window that eliminates contributions from alternatives III and IV in coincidence detection, thus restricting the two-photon interference to occur only between alternatives I and II. The experimental implementation is detailed in Methods section. We now analyze how a two-photon Laguerre-Gaussian (LG) mode $|l_s, p_s \rangle_s |l_i, p_i \rangle_i$ is transformed in each pathway.

\noindent{\bf{Alternative-I}}
\linebreak
To analyze the mode transformation in alternative-I, we first establish the relevant operators. A mirror reflection transformation can be represented by the operator  $\hat{M} = \sum_{l,p} e^{-i l \pi}|-l, p\rangle \langle l, p | $ \cite{karan2025sciadv}. Consequently, an even number of mirror reflections is equivalent to no change in the mode, that is $\hat{M}^{2n} = \mathbb{I}$, while an odd number of mirror reflections is equivalent to a single mirror reflection that is $\hat{M}^{2n+1} = \hat{M}$. The projection operator for an image rotator (IR) at rotation angle $\theta$ can be expressed as $\hat{\rm IR}\left(\theta \right)= \sum_{l,p} e^{-i l \left(\pi + 2\theta\right)}|-l,p\rangle \langle l,p|$ \cite{karan2025sciadv}.

Alternative-I corresponds to the case where both signal and idler photons take the longer path which is 80 cm longer than the path containing ${\rm IR}_s$ and ${\rm IR}_i$, as shown in the unfolded path alternative-I in Fig.~\ref{fig:si_path_alternatives}. In this configuration, both signal and idler undergo an even number of mirror transformations. Therefore, the projection operator for Alternative-I is $\mathbb{I}_s \otimes \mathbb{I}_i$, where $\mathbb{I}_{s(i)} $ is the identity projection operator for the signal (idler)  photon.

As a result, for a two-photon input mode $|l''_s, p''_s\rangle_s |l''_i, p''_i\rangle_i$ with signal and idler states of polarization along $\hat{\bm x}$ characterized by $\hat{\bm x}_s$ and $\hat{\bm x}_i$, respectively, the mode transformation can be written as  
\begin{equation}\label{eqn:si_alter_1}
|l''_s, p''_s\rangle_s |l''_i, p''_i\rangle_i \hat{\bm x}_s \otimes \hat{\bm x}_i \rightarrow  |l''_s, p''_s\rangle_s |l''_i, p''_i\rangle_i k_1 e^{i\left[ \phi_{s1} + \phi_{i1} - \omega_{s0}\left(\tau_{s1} + \tau_{i1} \right) \right]} \hat{\bm x}_s \otimes \hat{\bm x}_i,
\end{equation}
where $\otimes$ represents the tensor product between signal and idler photon states. $\omega_{s0}$ is the central frequency of the signal and idler photons for our case of degenerate SPDC. $k_1$ is the amplitude coefficient for this path alternative. 

\noindent{{\bf Alternative-II}}
\linebreak
Alternative-II corresponds to both signal and idler photons passing through the shorter arm containing ${\rm IR}_s$ and ${\rm IR}_i$, as shown in Fig.~\ref{fig:si_path_alternatives}. In this case, both photons undergo an odd number of mirror reflections and pass through ${\rm IR}_s$ and ${\rm IR}_i$ at angles $\theta_s$ and $\theta_i$, respectively. Consequently, the two-photon projection operator for this alternative is $\hat{\rm IR}_s\left(\theta_s\right)\hat{M}_s \otimes \hat{\rm IR}_i\left(\theta_i\right)\hat{M}_i$. We note that the transmitted states of polarization of photons change due to rotation of the image rotators, where $\hat{\bm \epsilon}_s\left(\theta_s\right)$ and  $\hat{\bm \epsilon}_i\left(\theta_i\right)$ represent the transmitted states of polarization for signal and idler photons passing through ${\rm IR}_s$ at rotation angle $\theta_s$  and  ${\rm IR}_i$ at rotation angle $\theta_i$, respectively \cite{karan2022ao, karan2024ao}. Therefore, for the two-photon input LG mode $|l''_s, p''_s\rangle_s |l''_i, p''_i\rangle_i \hat{\bm x}_s \otimes \hat{\bm x}_i$, the transformation is: 
\begin{equation}\label{eqn:si_alter_2}
|l''_s, p''_s\rangle_s |l''_i, p''_i\rangle_i \hat{\bm x}_s \otimes \hat{\bm x}_i \rightarrow  |l''_s, p''_s\rangle_s |l''_i, p''_i\rangle_i k_2 e^{i l_s 2\theta_s + i l_i 2\theta_i}e^{i\left[ \phi_{s2} + \phi_{i2} - \omega_{s0}\left(\tau_{s2} + \tau_{i2} \right) \right]} \hat{\bm \epsilon}_s\left(\theta_s\right) \otimes \hat{\bm \epsilon}_i\left(\theta_i\right),
\end{equation}
where $k_2$ is the amplitude coefficient for this path alternative. $\hat{\bm \epsilon}_s\left(\theta_s\right)$ can be expressed as  
\begin{equation} \label{eqn:si_transmitted_pol_s}
\hat{\bm \epsilon}_s\left( \theta_s\right)=  \cos\left[ \psi_s\left(\theta_s\right)\right] \hat{\bm x} +  \sin\left[ \psi_s\left(\theta_s\right)\right]e^{i\chi_s\left(\theta_s\right)} \hat{\bm y}.
\end{equation}
Here $\psi_s\left(\theta_s\right)$ and $\chi_s\left(\theta_s\right)$ are the azimuth and ellipticity angles of the transmitted state of polarization from ${\rm IR}_s$, respectively.  Similarly, $\hat{\bm \epsilon}_i(\theta_i)$ can be expressed as:
\begin{equation} \label{eqn:si_transmitted_pol_i}
\hat{\bm \epsilon}_i(\theta_i)=  \cos[\psi_i(\theta_i)] \hat{\bm x} + \sin[\psi_i(\theta_i)]e^{i\chi_i(\theta_i)} \hat{\bm y},
\end{equation}
where $\psi_i(\theta_i)$ and $\chi_i(\theta_i)$ are the corresponding angles for the idler photon passing through ${\rm IR}_i$.

Now, from Eqs.~(\ref{eqn:si_alter_1}) and (\ref{eqn:si_alter_2}), we can express the complete input mode transformation resulting from passage through the two-photon interferometer as:
\begin{align}\label{eqn:si_mode_transfrom_interfero}
|l''_s, p''_s\rangle_s |l''_i, p''_i\rangle_i \hat{\bm x}_s \otimes \hat{\bm x}_i \rightarrow  |l''_s, p''_s\rangle_s |l''_i, p''_i\rangle_i k_1 e^{i\left[ \phi_{s1} + \phi_{i1} - \omega_{s0}\left(\tau_{s1} + \tau_{i1} \right) \right]} \hat{\bm x}_s \otimes \hat{\bm x}_i \nonumber \\
+  |l''_s, p''_s\rangle_s |l''_i, p''_i\rangle_i k_2 e^{i l_s 2\theta_s + i l_i 2\theta_i}e^{i\left[ \phi_{s2} + \phi_{i2} - \omega_{s0}\left(\tau_{s2} + \tau_{i2} \right) \right]} \hat{\bm \epsilon}_s\left(\theta_s\right) \otimes \hat{\bm \epsilon}_i\left(\theta_i\right).
\end{align}
This transformation can be represented by the two-photon projection operator $\hat{P}^{(2)}$:
\begin{align} \label{eqn:si_projection_P}
&\hat{P}^{(2)} = \sum^{\infty}_{l''_s , l''_i = -\infty} \sum^{\infty}_{p''_s , p''_i = 0} e^{i \left[\gamma_1 + \phi_{s1} + \phi_{i1} - \omega_{s0}\left( \tau_{s1} + \tau_{i1}\right) \right]} \nonumber \\
&\times \left[|k_1|\hat{\bm x}_s \otimes \hat{\bm x}_i  + |k_2|e^{i\left( \delta + l_s 2\theta_s + l_i 2\theta_i\right)}  \hat{\bm \epsilon}_s\left(\theta_s\right) \otimes \hat{\bm \epsilon}_i\left(\theta_i\right) \right] |l''_s, p''_s\rangle_s |l''_i, p''_i\rangle_i {}_s \langle l''_s, p''_s| {}_i \langle l''_i, p''_i|,
\end{align}
where we define $k_1 = |k_1|e^{i \gamma_1}$, $k_2 = |k_2| e^{i \gamma_2}$ and $\delta = \left(\gamma_2 - \gamma_1 \right) + \phi_{s2} + \phi_{i2} - \phi_{s1} - \phi_{i1} - \omega_{s0} \left(\tau_{s2} + \tau_{i2} - \tau_{s1} -\tau_{i1} \right)$. We use this two-photon projection operator $\hat{P}^{(2)}$ at the coincidence detection plane to calculate the coincidence visibility for our detector configuration shown in Fig.~4(b) in the main text.

\section{Derivation of the coincidence visibility}\label{sec:si_coincidence_visibility}
In this section, we derive the expression for the coincidence visibility, which we measure experimentally to obtain the two-photon OAM spectrum. To accomplish this, we begin with the  two-photon input density matrix $\rho^{(2)}_{\rm in}$ in the Laguerre-Gaussian (LG) basis in the following form: 
\begin{equation} \label{eqn:si_rho_in}
%\rho^{(2)}_{\rm in} = \sum^{+N}_{l_s, l_i, l'_s, l'_i = -N}~ \sum^{\infty}_{p_s, p_i, p'_s,p'_i = 0} \langle C^{l_s,p_s}_{l_i, p_i} C^{* l'_s, p'_s}_{l'_i, p'_i} \rangle_{\rm e}|l_s, p_s\rangle_s |l_i, p_i\rangle_i  {}_s \langle l'_s, p'_s|{}_i \langle l'_i, p'_i |.
\end{equation}
The output density matrix from the two-photon interferometer shown in Fig.~1(b) in the main manuscript can be written as
\begin{equation} \label{eqn:si_rho_out}
\rho^{(2)}_{\rm out} = \hat{P}^{(2)} \rho^{(2)}_{\rm in} \hat{P}^{(2) \dag},
\end{equation}
where $\hat{P}^{(2)}$ is the two-photon projection operator at the coincidence detection plane. The expression for $\hat{P}^{(2)}$ is derived in Eq.~(\ref{eqn:si_projection_P}). Therefore, the angle-averaged coincidence detection probability as a function of $\theta_s$ and $\theta_i$ of ${\rm IR}_s$ and ${\rm IR}_i$, respectively can be written as 
\begin{align}\label{eqn:si_r_si}
R^{\delta}_{si} \left(\theta_s, \theta_i\right)&= \iint^{\infty}_{0}  \iint^{2 \pi}_{0} {}_s \langle \rho_s, \phi_s| {}_i \langle \rho_i, \phi_i|\rho^{(2)}_{\rm out} |\rho_s, \phi_s \rangle_s |\rho_i, \phi_i \rangle_i ~\rho_s \rho_i d\rho_s d\rho_i d\phi_s d\phi_i, \nonumber \\
&= \iint^{\infty}_{0}  \iint^{2 \pi}_{0} {}_s \langle \rho_s, \phi_s| {}_i \langle \rho_i, \phi_i| \hat{P}^{(2)} \rho^{(2)}_{\rm in} \hat{P}^{(2) \dag}  |\rho_s, \phi_s \rangle_s |\rho_i, \phi_i \rangle_i~ \rho_s \rho_i d\rho_s d\rho_i d\phi_s d\phi_i.
\end{align}
Substituting Eqs.~(\ref{eqn:si_projection_P}) and ~(\ref{eqn:si_rho_in}) into the above equation and using the relations $\hat{{\bm x}}_j \cdot \hat{{\bm x}}_j = \hat{{\bm y}}_j\cdot \hat{{\bm y}}_j =1 $ and $\hat{{\bm x}}_j \cdot \hat{{\bm y}}_j =0$, with $j= s,i$, we can express $R^{\delta}_{si} \left(\theta_s, \theta_i\right)$  as 
\begin{align}
&R^{\delta}_{si} \left(\theta_s, \theta_i\right) = \sum^{+N}_{l_s, l_i, l'_s, l'_i = -N}~ \sum^{\infty}_{p_s, p_i, p'_s,p'_i = 0} \langle C^{l_s,p_s}_{l_i, p_i} C^{* l'_s, p'_s}_{l'_i, p'_i} \rangle_{\rm e}\left[|k_1|^2 +|k_2|^2 e^{i~ 2\left(l_s - l'_s\right)\theta_s} e^{i~ 2\left(l_i - l'_i\right)\theta_i} \right. \nonumber \\
&\left. \qquad \qquad \qquad \qquad + |k_1||k_2|\cos\left[ \psi_s\left(\theta_s\right)\right] \cos\left[ \psi_i\left(\theta_i\right)\right]  \lbrace e^{-i\left(\delta + 2 l'_s \theta_s + 2l'_i \theta_i \right)} + e^{ i \left(\delta + 2 l_s \theta_s + 2l_i \theta_i\right) } \rbrace \right] \nonumber \\
&\times \iint^{\infty}_{0} \iint^{2 \pi}_{0} {}_s \langle \rho_s, \phi_s| {}_i \langle \rho_i, \phi_i| |l_s, p_s\rangle_s |l_i, p_i\rangle_i  {}_s \langle l'_s, p'_s|{}_i \langle l'_i, p'_i| |\rho_s, \phi_s \rangle_s |\rho_i, \phi_i \rangle_i ~ \rho_s \rho_i d\rho_s d\rho_i d\phi_s d\phi_i.
\end{align}
Using the orthogonality relations $\int^{\infty}_{0}\int^{2\pi}_{0} \left[LG^{|l'_s|}_{p'_s}\left(\rho_s\right) e^{-i l'_s \phi_s}\right]^{*}\left[LG^{|l_s|}_{p_s}\left(\rho_s\right) e^{-i l_s \phi_s}\right] \rho_s d\rho_s d \phi_s = \delta_{l'_s, l_s} \delta_{p'_s,p_s}$  and $\int^{\infty}_{0}\int^{2\pi}_{0} \left[LG^{|l'_i|}_{p'_i}\left(\rho_i\right) e^{-i l'_i \phi_i}\right]^{*}\left[LG^{|l_i|}_{p_i}\left(\rho_i\right) e^{-i l_i \phi_i}\right] \rho_i d\rho_i d \phi_i = \delta_{l'_i, l_i} \delta_{p'_i,p_i}$, we can simplify  $R^{\delta}_{si} \left(\theta_s, \theta_i\right)$ to:
\begin{align}
R^{\delta}_{si} \left(\theta_s, \theta_i\right) = \sum^{N}_{l_s, l_i =-N} & \sum^{\infty}_{p_s, p_i =0} \langle | C^{l_s,p_s}_{l_i, p_i}|^2\rangle_{\rm e} \left[ |k_1|^2 + |k_2|^2 \right. \nonumber \\
& \left. +  2|k_1||k_2| \cos\left[ \psi_s\left(\theta_s\right)\right] \cos\left[ \psi_i\left(\theta_i\right)\right]  \cos\left(\delta + 2l_s \theta_s + 2 l_i \theta_i\right) \right].
\end{align}
By definition, the two-photon OAM spectrum $P^{l_s}_{l_i}$ is the coincidence detection probability of the signal photon with OAM mode index $l_s$ and the idler photon with OAM mode index $l_i$, summed over all radial modes $p_s$ and $p_i$ present in the state for the same $l_s$ and $l_i$. That is, $P^{l_s}_{l_i}=\sum^{\infty}_{p_s, p_i =0} \langle | C^{l_s,p_s}_{l_i, p_i}|^2\rangle_{\rm e} $. Therefore, we can write $R^{\delta}_{si} \left(\theta_s, \theta_i\right)$ as 
\begin{align}\label{eqn:si_R_si_final}
&R^{\delta}_{si} \left(\theta_s, \theta_i\right)  =   |k_1|^2 + |k_2|^2 \nonumber \\
& +  2|k_1||k_2| \cos\left[ \psi_s\left(\theta_s\right)\right] \cos\left[ \psi_i\left(\theta_i\right)\right]  \sum^{N}_{l_s, l_i =-N}  P^{l_s}_{l_i} \cos\left(\delta + 2l_s \theta_s + 2 l_i \theta_i\right),
\end{align}
where we use the normalization condition ${\rm Tr}\left(\rho^{(2)}_{\rm in}\right)=1$, resulting in $\sum^{N}_{l_s, l_i =-N}  P^{l_s}_{l_i} =1$. 

Experimentally, measuring $R^{\delta}_{si} \left(\theta_s, \theta_i\right)$ as shown in Eq.~(\ref{eqn:si_R_si_final}) is challenging. The measurement suffers from accidental coincidences $R^{\delta}_n \left(\theta_s, \theta_i\right)$, primarily due to ambient multiphoton events, electronic noise in the coincidence detection circuitry etc. Additionally, changing the interferometer angles $\theta_s$ and $\theta_i$ for different measurements causes photon count fluctuations, which affects the coincidence detection probability. Moreover, pump laser intensity fluctuations during SPDC cause temporal variations in photon counts, and maintaining stable intensity throughout the entire measurement period is experimentally challenging. We define these fluctuations in the interferogram coincidence output as $f^{\delta}_{n} \left( \theta_s, \theta_i\right)$. Therefore, the measured coincidence detection probability can be expressed as:
\begin{align}
 \bar{R}^{\delta}_{si} \left(\theta_s, \theta_i\right) & =  f^{\delta}_{n} \left( \theta_s, \theta_i\right)  \Biggl[ R^{\delta}_n\left(\theta_s, \theta_i\right) + |k_1|^2  + |k_2|^2  \nonumber \\
&    +  2|k_1||k_2| \cos\left[ \psi_s\left(\theta_s\right)\right] \cos\left[ \psi_i\left(\theta_i\right)\right]  \sum^{N}_{l_s, l_i =-N}  P^{l_s}_{l_i} \cos\left(\delta + 2l_s \theta_s + 2 l_i \theta_i\right) \Biggr].
\end{align}
In order to overcome these issues in experimental measurements and therefore in reconstructing  $P^{l_s}_{l_i}$, we measure $\bar{R}^{\delta}_{si} \left(\theta_s, \theta_i\right)$  at two $\delta$  values, $\delta_c $ and $\delta_d$, and evaluate the coincidence visibility $\bar{V}_{si}$ as:
\begin{equation} \label{eqn:si_v_bar_si_common}
\bar{V}_{si}\left(\theta_s, \theta_i\right)= \dfrac{  \bar{R}^{\delta= \delta_c}_{si} \left(\theta_s, \theta_i\right) - \bar{R}^{\delta= \delta_d}_{si} \left(\theta_s, \theta_i\right)  }{ \bar{R}^{\delta= \delta_c}_{si} \left(\theta_s, \theta_i\right) + \bar{R}^{\delta= \delta_d}_{si} \left(\theta_s, \theta_i\right)  }.
\end{equation}
The explicit expression for ${V}_{si} \left(\theta_s, \theta_i\right)$  can be written as 
\begin{align} \label{eqn:si_v_si_gen}
&\bar{V}_{si}\left(\theta_s, \theta_i\right)= 2|k_1||k_2| \cos\left[ \psi_s\left(\theta_s\right)\right] \cos\left[ \psi_i\left(\theta_i\right)\right]  \nonumber \\
&\times  \sum^{N}_{l_s, l_i =-N}  P^{l_s}_{l_i} \left[ \left(\cos{\delta_c} -\cos{\delta_d}\right)\cos\left( 2l_s \theta_s + 2 l_i \theta_i\right)- \left(\sin{\delta_c} -\sin{\delta_d}\right)\sin\left( 2l_s \theta_s + 2 l_i \theta_i\right)\right] \Bigg/  \nonumber \\
&\Biggl[ 2 R^{\delta_c}_n\left(\theta_s, \theta_i\right) + 2|k_1|^2 + 2|k_2|^2 +  2|k_1||k_2| \cos\left[ \psi_s\left(\theta_s\right)\right] \cos\left[ \psi_i\left(\theta_i\right)\right]   \nonumber \\
&  \times  \sum^{N}_{l_s, l_i =-N}  P^{l_s}_{l_i} \left[ \left(\cos{\delta_c} + \cos{\delta_d}\right)\cos\left( 2l_s \theta_s + 2 l_i \theta_i\right)- \left(\sin{\delta_c} + \sin{\delta_d}\right)\sin\left( 2l_s \theta_s + 2 l_i \theta_i\right)\right]  \Biggr], 
\end{align}
where we assume that shot-to-shot accidental coincidence counts are negligible, that is, $R^{\delta_c}_n\left(\theta_s, \theta_i\right) \approx R^{\delta_d}_n\left(\theta_s, \theta_i\right)$ for each $\theta_s, \theta_i$ setting. Additionally, shot-to-shot coincidence fluctuations are also negligible, such that $ f^{\delta_c}_{n} \left( \theta_s, \theta_i\right)  \approx  f^{\delta_d}_{n} \left( \theta_s, \theta_i\right) $.

For the symmetric case of the two-photon OAM spectrum, that is,  $P^{l_s}_{l_i} = P^{-l_s}_{-l_i}$, we have $\cos\left(- 2l_s \theta_s - 2 l_i \theta_i\right) = \cos\left( 2l_s \theta_s + 2 l_i \theta_i\right)$ and $ \sin\left(- 2l_s \theta_s - 2 l_i \theta_i\right)  =    -\sin\left( 2l_s \theta_s + 2 l_i \theta_i\right) $. Next, substituting $\delta_c =0$ and $\delta_d =\pi$,  Eq.~(\ref{eqn:si_v_si_gen}) can be expressed as 
\begin{align} \label{eqn:si_v_si}
\bar{V}_{si}\left(\theta_s, \theta_i\right)=\dfrac{ 2|k_1||k_2| \cos\left[ \psi_s\left(\theta_s\right)\right] \cos\left[ \psi_i\left(\theta_i\right)\right]}{R^{\delta_c = 0}_n\left(\theta_s, \theta_i\right) + |k_1|^2 + |k_2|^2}  \sum^{N}_{l_s, l_i =-N}  P^{l_s}_{l_i}  \cos\left( 2l_s \theta_s + 2 l_i \theta_i\right).
\end{align}
Here we observe that $\bar{V}_{si}\left(\theta_s, \theta_i\right)$ does not depend on $f^{\delta}_{n} \left( \theta_s, \theta_i\right)$ and is therefore insensitive to coincidence fluctuations. However, the accidental count $R^{\delta_c}_n\left(\theta_s, \theta_i\right)$ remains in the denominator of Eq.~(\ref{eqn:si_v_si}). In our experimental conditions, this term has negligible effect, allowing us to approximate: $\frac{1}{R^{\delta_c = 0}_n\left(\theta_s, \theta_i\right) + |k_1|^2 + |k_2|^2} \approx \frac{1}{|k_1|^2 + |k_2|^2}$. Thus, $\bar{V}_{si}\left(\theta_s, \theta_i\right)$ becomes insensitive to accidental coincidence counts as well, and can then be expressed as:
\begin{align} \label{eqn:si_v_si_new}
\bar{V}_{si}\left(\theta_s, \theta_i\right)=\dfrac{ 2|k_1||k_2| \cos\left[ \psi_s\left(\theta_s\right)\right] \cos\left[ \psi_i\left(\theta_i\right)\right]}{|k_1|^2 + |k_2|^2}  \sum^{N}_{l_s, l_i =-N}  P^{l_s}_{l_i}  \cos\left( 2l_s \theta_s + 2 l_i \theta_i\right).
\end{align}
We note that the expression $\bar{V}_{si}\left(\theta_s, \theta_i\right)$ shown in Eq.~(\ref{eqn:si_v_si_new}) contains $\cos\left[\theta_s\right]\cos\left[\theta_i\right]$, which arise due to polarization change of the transmitted states from the ${\rm IR}_s$ and ${\rm IR}_i$ image rotators rotated at $\theta_s$ and $\theta_i$. We use the polarization-corrected coincidence visibility, defined as $V_{si}\left(\theta_s, \theta_i\right) = \bar{V}_{si}\left(\theta_s, \theta_i\right)/\left[ \cos\left[\theta_s\right]\cos\left[\theta_i\right] \right] $. With this correction, Eq.~(\ref{eqn:si_v_si_new}) becomes:
\begin{equation}
V_{si}\left(\theta_s, \theta_i\right)=\dfrac{ 2|k_1||k_2|}{|k_1|^2 + |k_2|^2}  \sum^{N}_{l_s, l_i =-N}  P^{l_s}_{l_i}  \cos\left( 2l_s \theta_s + 2 l_i \theta_i\right).
\end{equation}
We calibrate this interferometer and measure $\cos\left[\theta_s\right]$ and $\cos\left[\theta_i\right]$ experimentally, which correspond to the $x$-component of the transmitted states of polarization from ${\rm IR}_s$ and ${\rm IR}_i$, respectively. We define $|\cos\left[\theta_s\right]|^2  = \dfrac{|\hat{\bm \epsilon}\left(\theta_s\right) \cdot~ {\hat{\bm x}_s}|^2}{|\hat{\bm \epsilon}\left(\theta_s\right)|^2} = \dfrac{I_{s2x}\left(\theta_s\right)}{I^{\rm tot}_{s2}}$
and $|\cos\left[\theta_i\right]|^2  = \dfrac{|\hat{\bm \epsilon}\left(\theta_i\right) \cdot~ {\hat{\bm x}_i}|^2}{|\hat{\bm \epsilon}\left(\theta_i\right)|^2} = \dfrac{I_{i2x}\left(\theta_i\right)}{I^{\rm tot}_{i2}}$. 

Therefore, by measuring  $I_{s2x}\left(\theta_s\right)$, the intensity at the output of ${\rm IR}_s$ after a polarizer with its transmission axis along the ${\bm \hat{x}}$-direction at different $\theta_s$ and $I^{\rm tot}_{s2}$, the intensity without this polarizer, one can obtain $\cos\left[\theta_s\right]$. Similarly,  $\cos\left[\theta_i\right]$ for ${\rm IR}_i$ can be measured. From this calibration, we can obtain the polarization-corrected  coincidence visibility to measure the $P^{l_s}_{l_i}$ of any arbitrary two-photon general state.

\section{Non-symmetric two-photon OAM spectrum measurement}\label{sec:si_non_symmetric_spectrum_calculation}
While we present the measurement of the symmetric two-photon OAM spectrum in the main text, it is also possible to measure non-symmetric spectrum where $P^{l_s}_{l_i} \neq P^{-l_s}_{-l_i} $ using this detection technique. For non-symmetric OAM-spectrum measurements, two interferometric settings are insufficient. Instead, we require four such measurements. Here, we present how to measure the non-symmetric OAM spectrum via four-shot measurements of coincidence detection probabilities at $\delta =0 , \pi, \pi/2 ,$ and $3\pi/2$, whereas only two measurements at $\delta =0,$ and $\pi$ are required for the symmetric spectrum. 

Using Eq.~(\ref{eqn:si_v_si_new}) for the measured coincidence detection probability, we write the measured coincidence visibility for $\delta_c = 0$ and $\delta_d =\pi$ as: 
\begin{equation}\label{eqn:si_vsi_zero_pi}
\bar{V}^{(0,\pi)}_{si}\left(\theta_s, \theta_i\right)=\dfrac{ 2|k_1||k_2| \cos\left[ \psi_s\left(\theta_s\right)\right] \cos\left[ \psi_i\left(\theta_i\right)\right]}{|k_1|^2 + |k_2|^2}  \sum^{N}_{l_s, l_i =-N}  P^{l_s}_{l_i}  \cos\left( 2l_s \theta_s + 2 l_i \theta_i\right).
\end{equation}
Similarly, for $\delta_c = 3\pi/2$ and $\delta_d = \pi/2$, the measured coincidence visibility can be expressed as  
\begin{equation}\label{eqn:si_vsi_3pi_pi}
\bar{V}^{(3\pi/2, \pi/2)}_{si}\left(\theta_s, \theta_i\right)=\dfrac{ 2|k_1||k_2| \cos\left[ \psi_s\left(\theta_s\right)\right] \cos\left[ \psi_i\left(\theta_i\right)\right]}{|k_1|^2 + |k_2|^2}  \sum^{N}_{l_s, l_i =-N}  P^{l_s}_{l_i}  \sin\left( 2l_s \theta_s + 2 l_i \theta_i\right).
\end{equation}
We then apply polarization correction to the coincidence visibility, following the same procedure as for the symmetric spectrum presented in the main text. The polarization-corrected visibilities are defined as $V^{(0,\pi)}_{si}\left(\theta_s, \theta_i\right)= \dfrac{\bar{V}^{(0,\pi)}_{si}\left(\theta_s, \theta_i\right)}{ \cos\left[ \psi_s\left(\theta_s\right)\right] \cos\left[ \psi_i\left(\theta_i\right)\right]}$ and  $V^{(3\pi/2,\pi/2)}_{si}\left(\theta_s, \theta_i\right)= \dfrac{\bar{V}^{(3\pi/2,\pi/2)}_{si}\left(\theta_s, \theta_i\right)}{ \cos\left[ \psi_s\left(\theta_s\right)\right] \cos\left[ \psi_i\left(\theta_i\right)\right]}$.

With this correction, the polarization-corrected coincidence visibilities become: 
\begin{equation}\label{eqn:si_corrected_vsi_0_pi}
V^{(0,\pi)}_{si}\left(\theta_s, \theta_i\right)=\dfrac{ 2|k_1||k_2|}{|k_1|^2 + |k_2|^2}  \sum^{N}_{l_s, l_i =-N}  P^{l_s}_{l_i}  \cos\left( 2l_s \theta_s + 2 l_i \theta_i\right), 
\end{equation}
\begin{equation}\label{eqn:si_corrected_vsi_3pi_pi}
%V^{(3\pi/2, \pi/2)}_{si}\left(\theta_s, \theta_i\right)=\dfrac{ 2|k_1||k_2|}{|k_1|^2 + |k_2|^2}  \sum^{N}_{l_s, l_i =-N}  P^{l_s}_{l_i}  \sin\left( 2l_s \theta_s + 2 l_i \theta_i\right).
\end{equation}
From Eqs.~(\ref{eqn:si_corrected_vsi_0_pi}) and (\ref{eqn:si_corrected_vsi_3pi_pi}),  and through the two-dimensional Fourier transformations we obtain the measured two-photon OAM spectrum $\bar{P}^{l_s}_{l_i}$  as 
\begin{align} \label{eqn:si_bar_plsli_ftrans}
%\bar{P}^{l_s}_{l_i} & \equiv  \iint^{\pi/2}_{-\pi/2} \left[V^{(0,\pi)}_{si}\left(\theta_s, \theta_i\right) + i~ V^{(3\pi/2, \pi/2)}_{si}\left(\theta_s, \theta_i\right) \right] e^{-i \left(2l_s \theta_s + 2l_i \theta_i\right)}d\theta_s d\theta_i  \\
 &= \dfrac{ 2|k_1||k_2|}{|k_1|^2 + |k_2|^2} \sum^{N}_{l'_s, l'_i = -N} P^{l'_s}_{l'_i} \delta_{l_s,l'_s} \delta_{l_i,l'_i}, \nonumber \\
 &= \dfrac{ 2|k_1||k_2|}{|k_1|^2 + |k_2|^2}  P^{l_s}_{l_i} .
\end{align}
We note that the measured two-photon OAM spectrum $\bar{P}^{l_s}_{l_i}$ is not normalized. Using the normalization condition $ {\rm Tr}\left [\rho^{(2)}_{\rm in}\right] =1$, we obtain the normalized two-photon OAM spectrum  $P^{l_s}_{l_i}$ as
\begin{equation}
P^{l_s}_{l_i}= \dfrac{\bar{P}^{l_s}_{l_i}}{\sum^{N}_{l_s,l_i =-N} \bar{P}^{l_s}_{l_i} }.
\end{equation} 
Thus, we measure the non-symmetric $P^{l_s}_{l_i}$ by acquiring $\bar{V}^{(\delta_c, \delta_d)}_{si}\left(\theta_s, \theta_i\right)$ at four phase values $\delta = 0, \pi, \pi/2, 3\pi/2$ for each combination of $\theta_s$ and $\theta_i$. However, for the symmetric spectrum where $P^{l_s}_{l_i} = P^{-l_s}_{-l_i}$, only measurements at two phase settings $\delta = 0, \pi$ are sufficient. Therefore, without prior knowledge of whether the spectrum is symmetric or non-symmetric, we can measure $P^{l_s}_{l_i}$ through four-shot measurements.

\section{Effect of clipping in two-photon OAM spectrum measurement}\label{sec:si_effect_of_clipping}
\begin{figure}[t!]
\centering
\includegraphics[scale=1]{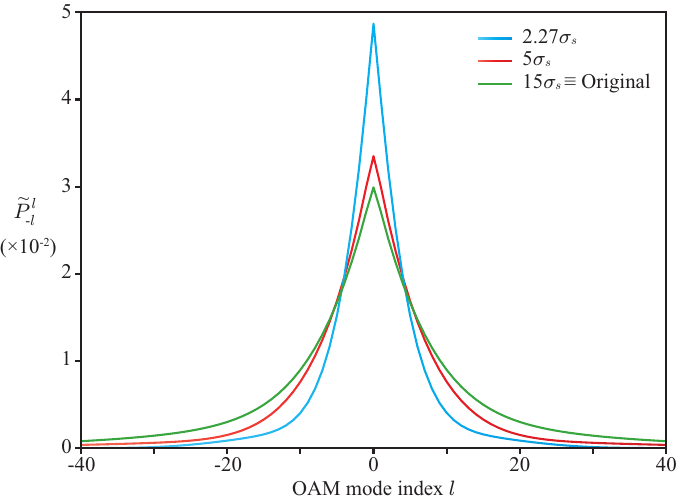}
\caption{{\bf Effect of collection aperture size on two-photon OAM mode detection.} Plot of $\widetilde{P}^{l}_{-l}$ as a function of OAM mode index $l$ for three different collection aperture radii $\rho_0$: $\rho_0 = 2.27\sigma_s$, $5\sigma_s$, and $15\sigma_s$. $\sigma_s$ is the radius of input signal or idler profile.}
\label{fig:si_effect_of_clipping}
\end{figure}
In the measurement of OAM spectrum, complete collection of the entire beam is crucial to obtain the actual spectrum. Any clipping of the incident state can lead to significant changes in the observed OAM spectrum.  Consequently, characterizing the extent of clipping is essential in experimental situations. 

To analyze this effect, we begin with the expression for the OAM Schmidt spectrum generated from SPDC, denoted as $P^{l}_{-l}$. This represents the antidiagonal element of the two-dimensional, two-photon OAM spectrum $P^{l_s}_{l_i}$. The expression for $P^{l}_{-l}$ generated from SPDC, without any post-generation clipping, is given by \cite{kulkarni2018pra}: 
\begin{align} \label{eqn:si_expression of Sl}
P^{l}_{-l}= \frac{1}{4 \pi^2} \int\!\!\!\!\int_{0}^{\infty}\rho_s \rho_i \Bigg|\int\!\!\!\!\int_{-\pi}^{\pi} V(\rho_s, \rho_i,\phi_s,\phi_i) \Phi(\rho_s, \rho_i,\phi_s,\phi_i) e^{-i l(\phi_s - \phi_i)} d \phi_s  d \phi_i \Bigg|^2  d\rho_s  d \rho_i .
\end{align}
However, in experiments, coupling a highly divergent SPDC field to a multimode fiber of finite core diameter and finite numerical aperture (NA) is technically demanding. Matching the beam diameter with the core diameter while maintaining the NA of the demagnified beam below the fiber's NA presents significant challenges. As a result, radial clipping of the beam is often unavoidable. To model this effect theoretically, we consider that the signal intensity profile has a waist size of $\sigma_s$, with the same parameter applying to the idler. The collection aperture radius is defined as $\rho_0$. With this collection aperture size, the expression for the clipped two-photon OAM spectrum for $l_s = -l_i$, defined by $\widetilde{P}^{l}_{-l}$ becomes
\begin{align}\label{eqn:si_expression Sl clipped}
\widetilde{P}^{l}_{-l}=\frac{1}{4 \pi^2} \int\!\!\!\!\int_{0}^{\rho_0}\rho_s \rho_i \Bigg|\int\!\!\!\!\int_{-\pi}^{\pi} V(\rho_s, \rho_i,\phi_s,\phi_i) \Phi(\rho_s, \rho_i,\phi_s,\phi_i) e^{-i l(\phi_s - \phi_i)} d \phi_s  d \phi_i \Bigg|^2  d\rho_s  d \rho_i ,
\end{align}
where when $\rho_0$ is sufficiently larger than $\sigma_s$, we have $\widetilde{P}^{l}_{-l} = P^{l}_{-l}$. Figure~\ref{fig:si_effect_of_clipping} presents numerical plots of $\widetilde{P}^{l}_{-l}$ as a function of OAM mode index for $\rho_0 = 2.27\sigma_s$, $5\sigma_s$, and $15\sigma_s$. $\widetilde{P}^{l}_{-l}$ at $\rho_0 = 15\sigma_s$ is essentially the original unclipped $P^{l}_{-l}$, corresponding to a sufficiently large aperture. We observe that as $\rho_0$ decreases, the width of $\widetilde{P}^{l}_{-l}$ decreases. Physically, clipping the higher-order wave vectors is equivalent to filtering out higher-order OAM modes from detection.

We have characterized our experimental setup, shown in Fig.~4(b) in the main text, which uses a lens combination to couple the output interferogram to a multimode fiber with a core diameter of $62~\mu$m and numerical aperture of $0.275$ before detection at the SPCM. For this configuration, we find that $\rho_0 = 2.27 \sigma_s$. Although this captures more than $96\%$ of the total optical power, $\widetilde{P}^{l}_{-l}$  is significantly truncated relative to the original $P^{l}_{-l}$, as demonstrated in the simulation shown in Fig.~\ref{fig:si_effect_of_clipping}. 

Therefore, in our theoretical description of the two-photon OAM spectrum generation from SPDC and in the calculation of $ P^{l_s}_{l_i}$, we incorporate the experimental constraint by setting $ \rho_0 = 2.27 \sigma_s $. Thus, the generated two-photon OAM spectrum from SPDC is inherently truncated, matching what is detected by our OAM spectrum measurement technique.

\end{document}